\documentclass{IEEEoj}
\usepackage{cite}
\usepackage{amsmath,amssymb,amsfonts}
\usepackage{amsmath,amssymb,amsfonts}
\usepackage{algorithm}
\usepackage{algpseudocode}
\usepackage{graphicx,color}
\usepackage{textcomp}
\usepackage{acronym}

\usepackage{array}
\newcolumntype{C}[1]{>{\centering\arraybackslash}m{#1}}
\newcolumntype{P}[1]{>{\arraybackslash}m{#1}}
\usepackage{multirow}
\usepackage{dblfloatfix}

\acrodef{5G CN}[5G CN]{5G Core Network}

\acrodef{AGV}[AGV]{automated guided vehicle}
\acrodef{AI/ML}[AI/ML]{Artificial Intelligence/Machine Learning}
\acrodef{AI}[AI]{Artificial Intelligence}
\acrodef{AWS}[AWS]{Amazon Web Services}
\acrodef{ACPI}[ACPI]{Advanced Configuration and Power Interface}
\acrodef{AE}[AE]{Autoencoder}
\acrodef{API}[API]{application programming interface}

\acrodef{B5G}[B5G]{Beyond fifth-generation}

\acrodef{CNCF}[CNCF]{Cloud Native Computing Foundation}
\acrodef{CaaS}[CaaS]{Containers as a Service}
\acrodef{CI/CD}[CI/CD]{Continuous Integration/Continuous Deployment}
\acrodef{CUE}[CUE]{Carbon Usage Effectiveness}
\acrodef{CLI}[CLI]{command-line interface}
\acrodef{CDF}[CDF]{cumulative distribution function}

\acrodef{DC}[DC]{Data Center}
\acrodefplural{DC}{Data Centers} 
\acrodef{DL}[DL]{Downlink}

\acrodef{eMBB}[eMBB]{enhanced Mobile Broadband}
\acrodef{eBPF}[eBPF]{extended Berkeley Packet Filter}
\acrodef{EEGA}[EEGA]{Energy Efficient Genetic Algorithm}
\acrodef{XAI}[XAI]{eXplainable AI}

\acrodef{FC}[FC]{Fully Connected}
\acrodef{GAN}[GAN]{Generative
Adversarial Network}
\acrodef{GA}[GA]{Genetic Algorithm}
\acrodef{Gen-AI}[Gen-AI]{Generative AI}
\acrodefplural{GA}{Genetic Algorithms}
\acrodef{gNB}[gNB]{Next Generation Node B}

\acrodef{IoT}[IoT]{Internet of Things}
\acrodef{IaaS}[IaaS]{Infrastructure as a Service}
\acrodef{ILP}[ILP]{Integer Linear Programming}

\acrodef{json}[json]{JavaScript Object Notation}

\acrodef{K8s}[K8s]{Kubernetes}
\acrodef{Kepler}[Kepler]{Kubernetes Efficient Power Level Exporter}
\acrodef{KPI}[KPI]{Key Performance Indicator}
\acrodef{KL}[KL]{Kullback-Leibler}
\acrodef{KS Test}[KS Test]{Kolmogorov-Smirnov test}
\acrodef{KPMMON}[KPMMON]{key performance metric monitoring}

\acrodef{LAN}[LAN]{Local Area Network}
\acrodef{LOF}[LOF]{Local Outlier Factor}
\acrodef{LSTM}[LSTM]{Long Short Term Memory}

\acrodef{mMTC}[mMTC]{massive Machine-Type Communications}
\acrodef{ML}[ML]{Machine Learning}
\acrodef{MQTT}[MQTT]{Message Queuing Telemetry Transport}
\acrodef{MILP}[MILP]{Mixed Integer Linear Programming}
\acrodef{MNO}[MNO]{Mobile Network Operator}
\acrodef{MLP}[MLP]{Multi-Layer Perceptron}
\acrodef{MSE}[MSE]{Mean Squared Error}
\acrodef{MRTT}[MRTT]{Model Retrain Trigger Time}
\acrodef{MRPT}[MRPT]{Model Re-Placement Time}
\acrodef{Non-RT RIC}[Non-RT RIC]{Non-real time RIC}
\acrodef{Near-RT RIC}[Near-RT RIC]{Near-real time RIC}
\acrodef{NS}[NS]{Network Slicing}

\acrodef{OAI}[OAI]{OpenAirInteraface}
\acrodef{O-RAN}[O-RAN]{Operator Defined Open and Intelligent Radio Access Networks}
\acrodef{OSC}[OSC]{O-RAN software community}
\acrodef{OpEx}[OpEx]{Operational Expenditure}
\acrodef{O-CU}[O-CU]{Central Unit}
\acrodef{O-DU}[O-DU]{Distributed Unit}
\acrodef{O-RU}[O-RU]{Radio Unit}

\acrodef{PaaS}[PaaS]{Platform as a Service}
\acrodef{PUE}[PUE]{Power Utilization Effectiveness}
\acrodef{PMF}[PMF]{Probability Mass Function}

\acrodef{QoS}[QoS]{Quality of Service}
\acrodef{QP}[QP]{QoS prediction}

\acrodef{RL}[RL]{Reinforcement Learning}
\acrodef{RAPL}[RAPL]{Running Average Power Limit}
\acrodef{RIC}[RIC]{RAN Intelligent Controllers}
\acrodef{RAN}[RAN]{Radio Access Network}
\acrodef{ReLU}[ReLU]{rectified linear unit}
\acrodef{RMSE}[RMSE]{Root Mean Square Error}
\acrodef{REST}[REST]{Representational State Transfer}

\acrodef{SaaS}[SaaS]{Software as a Service}
\acrodef{SLA}[SLA]{Service Level Agreements}
\acrodef{SMO}[SMO]{Service Management and Orchestration}
\acrodef{TIP}[TIP]{Telecom Infra Project}
\acrodef{TGEN}[TGEN]{Colosseum traffic generator}
\acrodef{uRLLC}[uRLLC]{ultra-Reliable Low Latency Communications}
\acrodef{UE}[UE]{User Equipment}
\acrodef{UL}[UL]{Uplink}

\acrodef{VAE}[VAE]{Variational Autoencoder}
\acrodef{VWall}[VWall]{Virtual Wall}
\acrodef{VNF}[VNF]{Virtual Network Function}
\acrodefplural{VNF}{Virtual Network Functions} 
\acrodef{VM}[VM]{Virtual Machine}
\acrodefplural{VNF}{Virtual Machines} 

\acrodef{WUE}[WUE]{Water Usage Effectiveness}
\acrodef{WLAN}[WLAN]{Wireless Local Area Network}
\acrodef{WSAN}[WSAN]{Wireless Sensor and Actuator Network}

\def\BibTeX{{\rm B\kern-.05em{\sc i\kern-.025em b}\kern-.08em
    T\kern-.1667em\lower.7ex\hbox{E}\kern-.125emX}}
\AtBeginDocument{\definecolor{ojcolor}{cmyk}{0.93,0.59,0.15,0.02}}
\def\OJlogo{\vspace{-4pt}\hskip-4pt\includegraphics[height=18pt]{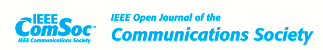}}
\begin{document}
\receiveddate{XX Month, XXXX}
\reviseddate{XX Month, XXXX}
\accepteddate{XX Month, XXXX}
\publisheddate{XX Month, XXXX}
\currentdate{11 January, 2024}
\doiinfo{OJCOMS.2024.011100}

\title{Generative-AI for AI/ML Model Adaptive Retraining in Beyond 5G Networks}

\author{Venkateswarlu~Gudepu\IEEEauthorrefmark{1} \IEEEmembership{(Graduate Student Member, IEEE)}, Bhargav~Chirumamilla\IEEEauthorrefmark{1}, Venkatarami~Reddy~Chintapalli\IEEEauthorrefmark{2}\IEEEmembership{(Graduate Student Member, IEEE)}, Piero~Castoldi\IEEEauthorrefmark{3}\IEEEmembership{(Senior Member, IEEE)}, Luca~Valcarenghi\IEEEauthorrefmark{3}\IEEEmembership{(Senior Member, IEEE)}, Bheemarjuna~Reddy~Tamma\IEEEauthorrefmark{4}\IEEEmembership{(Senior Member, IEEE)} AND Koteswararao~Kondepu\IEEEauthorrefmark{1}
\IEEEmembership{(Senior Member, IEEE)}}
\affil{Indian Institute of Technology Dharwad, Dharwad, Karnataka, India.}
\affil{National Institute of Technology Calicut, Calicut, India.}
\affil{Scuola Superiore Sant'Anna, Via Moruzzi 1, 56124 Pisa, Italy.}
\affil{Indian Institute of Technology Hyderabad, India.}
\corresp{CORRESPONDING AUTHOR: Venkateswarlu~Gudepu (e-mail: 212011003@iitdh.ac.in).}
\authornote{This study is partly supported through a DST SERB Startup Research Grant (SRG-2021-001522). This study has been sponsored partly by the KDT-JU project Collaborative edge-cLoud Continuum and Embedded AI for a Visionary Industry of thE futuRe (CLEVER) (grant agreement no. 101097560).
KDT-JU receives funding from the Horizon Europe Research Framework and the National Authorities.}
\markboth{Generative-AI for AI/ML Model Adaptive Retraining in Beyond 5G Networks}{Author \textit{et al.}}

\begin{abstract}
Beyond fifth-generation (B5G) networks aim to support high data rates, low-latency applications, and massive machine communications. 
\ac{AI/ML} can help to improve B5G network performance and efficiency.
However, dynamic service demands of B5G use cases cause \ac{AI/ML} model performance degradation, resulting in \ac{SLA} violations, over- or under-provisioning of resources, etc.
Retraining is essential to address the performance degradation of the \ac{AI/ML} models. 
Existing threshold and periodic retraining approaches have potential disadvantages, such as \ac{SLA} violations and inefficient resource utilization for setting a threshold parameter in a dynamic environment.
This paper proposes a novel approach that predicts when to retrain \ac{AI/ML} models using Generative Artificial Intelligence.
The proposed predictive approach is evaluated for a Quality of Service Prediction use case on the Open Radio Access Network (O-RAN) Software Community platform and compared to the predictive approach based on the classifier and a threshold approach.
Also, a real-time dataset from the Colosseum testbed is considered to evaluate Network Slicing (NS) use case with the proposed predictive approach.
The results show that the proposed predictive approach outperforms both the classifier-based predictive and threshold approaches.
\end{abstract}


\maketitle

\section{Introduction}\label{sec:intro}

\IEEEPARstart{T}{he} \ac{B5G} networks are bringing transformations in the next-generation networks by supporting a range of use cases such as \ac{mMTC}, \ac{uRLLC}, and \ac{eMBB}. 
However, \acp{MNO} recognizes that there is a strong need for network intelligence to address the complexity of \ac{B5G} networks effectively and meet the growing service demands. 
Artificial Intelligence/Machine Learning (AI/ML) techniques offer a promising solution as they can effectively handle complex network architectures and make intelligent decisions (i.e., resource allocation based on the predicted user traffic)~\cite{bartsiokas2022ml,khan2023ai}, which makes them suitable for \ac{B5G} networks.

\ac{O-RAN}, \ac{TIP}, and Open RAN Policy Coalition, and many others are actively working towards enabling the \emph{intelligence} in \ac{B5G} networks. 
Here, the \ac{O-RAN} Alliance architecture enables \emph{intelligence} through two logical \acp{RIC}: \ac{Non-RT RIC} and \ac{Near-RT RIC}~\cite{oranwg2}. 
The \ac{Non-RT RIC} operates use cases with a granularity of at least $1\ s$, while the \ac{Near-RT RIC} operates use cases on a timescale between $10\ ms$ and $1\ s$.

Utilizing \ac{AI/ML} models enhances the performance of \ac{B5G} networks, but it comes with several challenges~\cite{aryal2023open}. One significant challenge is maintaining performance consistency with \ac{AI/ML} model --- which is crucial for decision-making in network management use cases. 
The dynamic changes in incoming user data traffic in \ac{B5G} networks contribute to model performance degradation, thus impacting \ac{AI/ML}-based applications~\cite{manias2023model}. 
This issue is especially critical, as it can lead to false insights, inaccurate decision-making, and potentially severe service interruptions. 
Moreover, it results in inefficient resource allocation and utilization~\cite{gudepu2023exploiting}, where over/under-provisioning can lead to network performance degradation and service disruptions.

The above factors influence the pre-defined \acp{SLA} between service providers and users~\cite{slicem}. For example, in an \ac{AGV} use case, the \ac{SLA} defines a predicted throughput of $80\ Mbps$. 
However, when an \ac{AI/ML} model performance degradation occurs, the model may not predict the expected SLA throughput, which leads to violations of pre-defined \acp{SLA}~\cite{5gverticals}. 
Therefore, addressing the model performance degradation is crucial in \ac{B5G}. 

To ensure the \ac{AI/ML} model performance --- the model retraining (or updating) with newly arrived user data is an effective strategy. 
In~\cite{al2020prediction}, a threshold approach is proposed to trigger model retraining. 
This approach continuously monitors key performance metrics such as \emph{accuracy}, \emph{precision}, \emph{recall}, or \emph{F1-score} and initiates retraining when these metrics fall below or exceed a pre-defined threshold.~\cite{mekrachecombining} triggers retraining whenever the \ac{AE} reconstruction error violates the pre-defined threshold. 
However, determining the appropriate threshold value poses significant challenges --- too low may result in excessive computational costs due to frequent retraining --- too high may lead to poor model performance and violate \acp{SLA}. 
In~\cite{samdanis2023ai}, the models are updated at regular intervals regardless of changes in the user traffic or performance of the deployed \ac{AI/ML} model. 

The \ac{AI/ML} model retraining is predicted (namely predictive approach) by determining changes in user traffic through an unsupervised classifier --- \ac{LOF}~\cite{Gudepu2023AdaptiveRO}. This approach minimizes \ac{SLA} violations and optimizes resource utilization. However, classifier based predictive approach~\cite{Gudepu2023AdaptiveRO} has certain limitations, such as --- $(i)$ fails to determine changes in user traffic whenever the changes are not continuous; 
$(ii)$ may not be suitable for the incoming user traffic with high variance and can lead to frequent retraining; and 
$(iii)$ the method used to determine the number of consecutive windows (or data chunks) to trigger model retraining might vary depending on the specific application being considered.

To address the limitations of statistical measures within the threshold approach~\cite{al2020prediction, mekrachecombining} and unsupervised classifiers employed in the predictive approach~\cite{Gudepu2023AdaptiveRO} for \ac{AI/ML} model retraining, this paper introduces a novel predictive approach by utilizing the capabilities of \ac{Gen-AI}~\cite{ganretrain,karapantelakis2023generative}. \ac{Gen-AI} capitalizes advanced features to effectively capture and model the underlying distribution of user traffic, unlike traditional statistical measures that have limitations in handling complex and dynamic data distributions. 
\ac{Gen-AI} can adapt to complex user traffic patterns in dynamic environments by enabling the dynamic nature of user behavior within \ac{B5G} networks.

\ac{Gen-AI} enables more realistic simulations of diverse network scenarios~\cite{navidan2021generative}, aiding in predicting \ac{AI/ML} model retraining to align with evolving user traffic patterns in dynamic \ac{B5G} environments, thus enhancing predictive capabilities and reducing \ac{SLA} violations.

Incorporating \ac{Gen-AI} is significant for mitigating \ac{SLA} violations and improving computational resource efficiency by generating synthetic data resembling observed traffic patterns.

The main contributions of this paper are summarized as follows: 
\begin{itemize}
    \item A novel approach to predict the \ac{AI/ML} model retraining using \ac{Gen-AI} --- \acp{VAE} and \acp{GAN}.
    \item The proposed predictive approach is evaluated for both the QoS prediction use case over the O-RAN software community (OSC) platform~\cite{OSC} and a real-time slicing dataset from the Colosseum~\cite{bonati2021intelligence}.
    \item Discussion on various challenges and complexities involved in leveraging \ac{Gen-AI} for \ac{B5G} networks. 
\end{itemize}

\section{Background}
\label{sec:backgroundwork}

Generative AI --- \ac{VAE} and \ac{GAN}, is a broad category of \ac{AI/ML} techniques that focus on generating new user traffic data that closely resembles observed (or trained) data. 
\acp{VAE} generates data by capturing the observed data that underlie the distribution and sampling from a learned latent space. 
On the other hand, \acp{GAN} uses a combination of the \textit{generator} and \textit{discriminator} networks to generate new user traffic data. 
The \textit{generator} network generates synthetic data, also known as ``fake data'' or ``generated data,'' to resemble the distribution of the observed data. Whereas, the \textit{discriminator} network distinguishes between the observed and the generated data. 

The detailed descriptions of both \ac{VAE} and \ac{GAN} as follows: 

\subsection{\acp{VAE}}

\begin{figure}[h]
    \centering
    \includegraphics[width=1.0\linewidth]{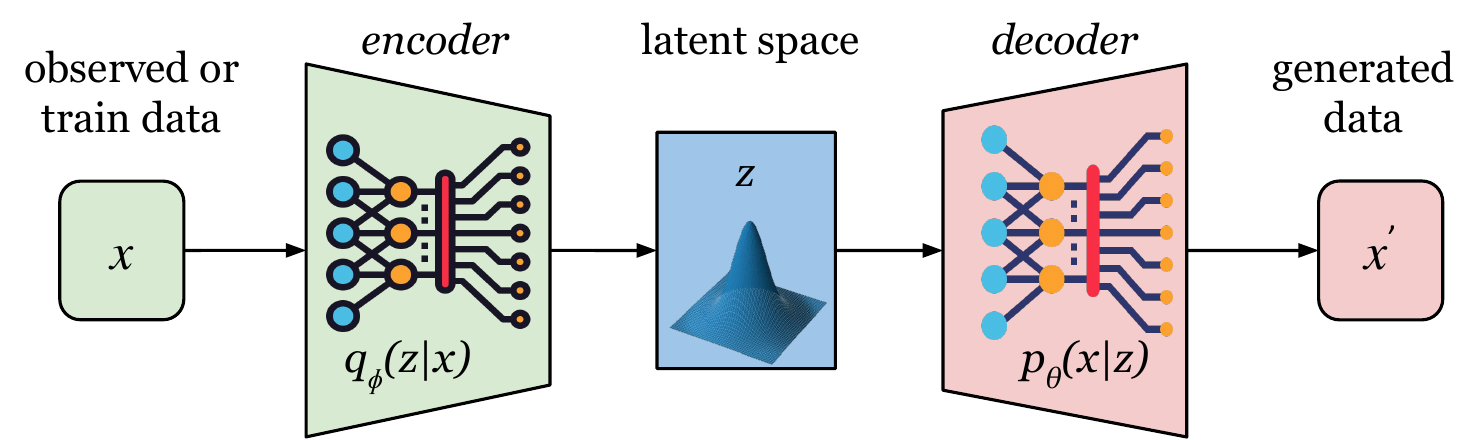}
    \caption{VAE architecture}
    \label{fig:VAE}
\end{figure}

Figure~\ref{fig:VAE} shows the \ac{VAE} architecture consists of two neural networks: an \emph{encoder} and a \emph{decoder}. 
The encoder takes user traffic data (i.e., \(x\)), and transforms into a latent space representation (i.e., \(z\)) --- describes the probability distribution of every input value and approximate new user traffic data~\cite{bank2023autoencoders}. 
The \textit{encoder} is represented as \(q_\phi(z|x)\) and parameterized by \(\phi\) to provide the mean and variance of a $Normal\ distribution$ from which the latent variable \(z\) is sampled. 
The \emph{decoder} is denoted as \(p_\theta(x|z)\) and parameterized by \(\theta\), takes the latent variable \(z\) as input and generates the parameters of the distribution \(x\). 
The primary objective of the \ac{VAE} is twofold:  (i) to minimize the reconstruction error of the user traffic data; and (ii) to minimize the \ac{KL} divergence~\cite{raiber2017kullback} between the observed data distribution \(p(z)\) and the newly generated data distribution (i.e., encoder) \(q_\phi(z|x)\) as in Equation~\eqref{Eq:VAE_loss_function}.
\begin{equation}
    \mathrm{V}(\theta, \phi)=\mathbb{E}_{q_{\phi}(z \mid x)}\left[\log p_{\theta}(x \mid z)\right]-D_{K L}\left(q_{\phi}(z \mid x) \| p_{\theta}(z)\right)
    \label{Eq:VAE_loss_function}
\end{equation}
The \ac{VAE} can generate new user traffic data upon training, where latent variable is \(z_i \sim p(z)\) and a new user traffic data sample is  \(x_i \sim p(x|z)\).

\subsection{\ac{GAN}}

\begin{figure}[htb!]
\centering
\includegraphics[width=1.0\linewidth]{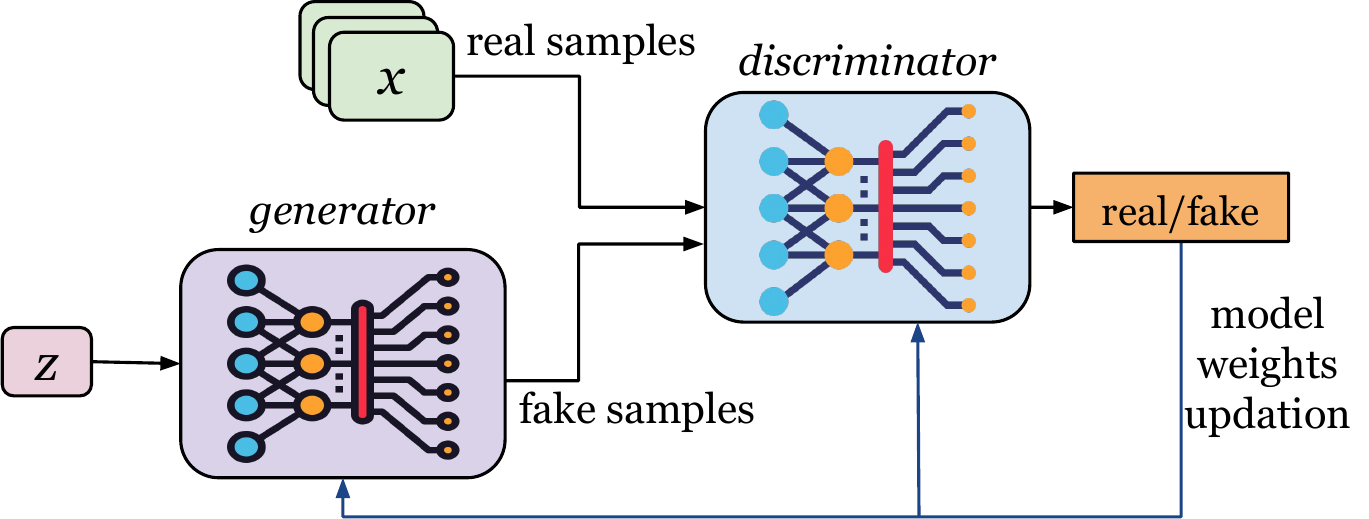}
\caption{GAN architecture.}
\label{fig:gan_topology}
\end{figure}
Figure~\ref{fig:gan_topology} shows the \ac{GAN} architecture, which consists of two neural networks: the \emph{generator} and the \emph{discriminator}.
In the \ac{GAN} architecture, adversarial training is employed, where a generator competes with a discriminator to produce realistic data. 
The \acp{GAN}  have demonstrated in capturing complex distributions and generating user traffic closer to actual traffic~\cite{ayanoglu2022machine}. 
The \acp{GAN} does not make assumptions about a specific distribution for latent space representation as in \acp{VAE}. Thus, the \acp{GAN} offer increased flexibility in modeling various user traffic patterns.  

The generator network takes a random \emph{noise} vector $z$ as input and attempts to create fake data, which resembles the input data traffic during training. 
In contrast, the discriminator attempts to classify the generated data from the observed data accurately. 
Loss convergence of the generator and discriminator completes the training period. 
Essentially, both the generator and discriminator networks are jointly involved in a 2 -- player min-max game, as shown in Equation~\ref{Eq:GAN}, until the discriminator fails to distinguish between the observed and generated data~\cite{kaloxylos2021ai}.

\begin{align}
\min_{\theta_{G}} \max_{\theta_{D}} L(G, D) &= \min_{G} \max_{D} \mathbb{E}_{x\sim p_{\text{data}}(x)}[\log{D(x)}] \nonumber \\
&\quad+ \mathbb{E}_{z\sim p_{\text{z}}(z)}[\log{1 - D(G(z))}] 
\label{Eq:GAN}
\end{align}

Where $x$ is the user input data, $p_{\text{data}}$ is the observed data distribution, $p_{\text{z}}(z)$ is the noise distribution, $log(D(x))$ is the predicted output of the discriminator for $x$, and $log(D(G(z))$ is the output of the discriminator on the \ac{GAN} generated data $G(z)$. 
The aim is to maximize the ability of the discriminator to identify observed data from generated data (i.e., $\max\limits_{D}$). 
In contrast, the generator part of the Equation tries to minimize the discriminator's ability to classify observed and generated data correctly (i.e., $\min\limits_{G}$). 

The applications of \ac{Gen-AI} in the \ac{RAN} include generating synthetic data, resource optimization, network planning, and \ac{QoS} management ~\cite{ganretrain,karapantelakis2023generative,kaloxylos2021ai}.
However, the proposed predictive approach leverages the \ac{Gen-AI} to determine when to retrain an \ac{AI/ML} model, which is detailed in the section~\ref{sec:systemmodelandproposedapproach}.

\begin{figure*}[h]
\centering
\includegraphics[width=1.0\linewidth]{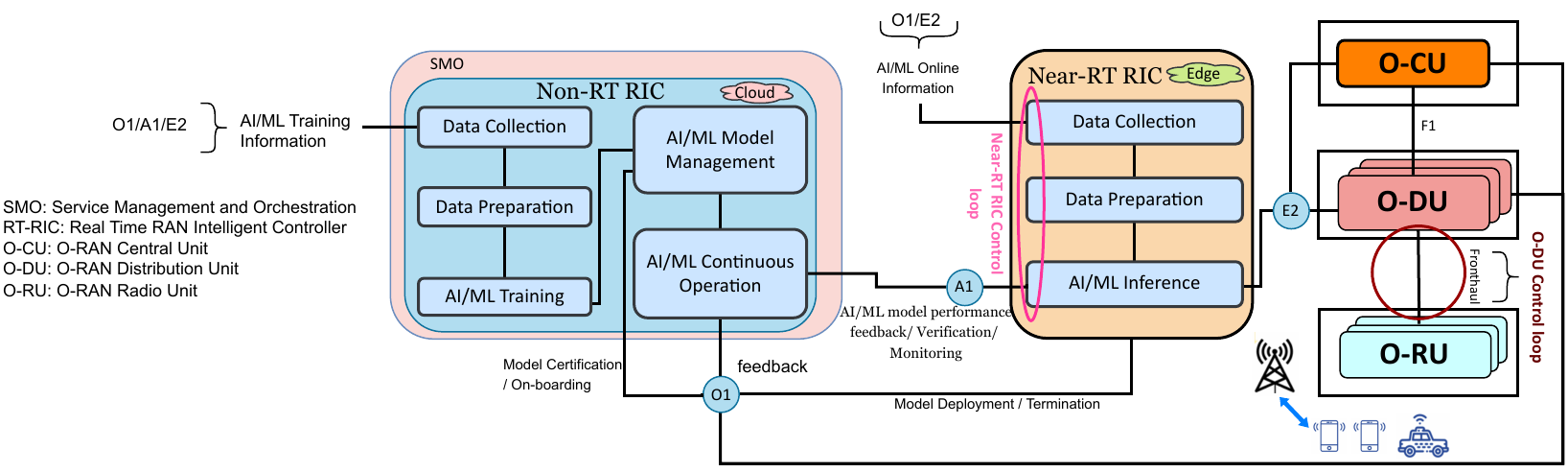}
\caption{System model.}
\label{fig:arch_eu}
\end{figure*}
\section{System Model and Proposed Approach}\label{sec:systemmodelandproposedapproach}

This section describes the system model and the proposed approach to predict when to retrain an AI/ML model. 

\subsection{System Model}
Figure~\ref{fig:arch_eu} shows the O-RAN architecture in which two \ac{RAN} \acp{RIC} are defined: the Near-RT and the Non-RT~\cite{oranwg2} to employ the \emph{intelligence}. 
These \acp{RIC} enable \ac{B5G} network autonomous optimization by operating at different timescales depending on the position of the \ac{AI/ML} model inference~\cite{chintapalli2022wip}.

The O-RAN architecture provides various interfaces, including O1, A1, and E2, to enable data collection and communication among the RAN components --- \ac{O-CU}, \ac{O-DU}, and \ac{O-RU}. 
The \emph{O1-interface} obtains the user traffic data from all the components (\ac{O-CU}, \ac{O-DU}, and \ac{O-RU} and model deployment/termination information from Non-RT to Near-RT RIC. 
The \emph{A1-interface} publishes policy-based guidance, \ac{AI/ML} performance feedback, verification, and monitoring information between Non-RT and Near-RT RIC. 
The \emph{E2-interface} is to control the \ac{RAN} functions through E2-control messages.

The \ac{Non-RT RIC} is part of the \ac{SMO} and facilitates radio resource management and policy optimization within the \ac{RAN}. 
Non-RT RIC consists of AI/ML model management to manage data collection, preparation, model training, and deploying AI/ML inference as an x/rApp. 
Simultaneously, \ac{AI/ML} continuous operation monitors deployed x/rApp performance and through the A1 interface. 
The Near-RT RIC module operates with granularity between $10\ ms$ and $1\ s$, enabling near-real-time optimization, control, and data monitoring of O-CU and O-DU nodes. 
\ac{Near-RT RIC} components include \ac{AI/ML} inference in the form of xApp, data collection and preparation to monitor the \ac{RAN} performance. 
The details of the proposed approach are in the following section.

\subsection{Proposed Approach}

The proposed approach predicts when to retrain the \ac{AI/ML} model by exploiting \ac{Gen-AI} --- \acp{VAE} and \acp{GAN} --- to generate data that are close to the observed and then performs the \ac{KS Test}~\cite{marsaglia2003evaluating} over both the generated and incoming user traffic data to trigger retrain. Here, the KS Test is used to determine whether the incoming user traffic follows the observed data distribution or not. 
Algorithms~\ref{alg_VAE} and~\ref{alg_GAN} depict the proposed predictive approach implemented using both \acp{VAE} and \acp{GAN} respectively. Table~\ref{tab:acronym} reports the definitions of the parameters/variables that are used.

\begin{table*}[!t]
\small
    \caption{Description of variables used in the algorithms.}
    \label{tab:acronym}
    \begin{tabular}{|p{0.179\columnwidth}|p{1.79\columnwidth}|}
  \hline
\textbf{Acronym}  & \textbf{Referring to / Definition}       \\ \hline

\textit{${ds}$} & Incoming data stream\\ \hline

\textit{$\mathcal{DE}$} & Decoder neural network architecture of \ac{VAE}\\ \hline

\textit{$P_{VAE-kstest}$} & Value determined by Kolmogorov-Smirnov Test (KS Test) between the generated and observed data during the training of VAE \\ 
\hline

\textit{$WS$} & Window Size \\ 
\hline

\textit {Noise} & A source of variability or randomness injected into the \ac{GAN} model to generate diverse and realistic data \newline The proposed work employed Gaussian noise throughout the study\\ 
\hline

\textit {$z$} & Samples from the distribution considered in $Noise$\\ 
\hline

\textit {$gd_{\text {data }}$} & Generated (or new) data from Decoder $\mathcal{DE}$\\ 
\hline

\textit{$\mathcal{DE}({z})$} & Decoder $\mathcal{DE}$, taking a vector $z_i$ of size $WS$ as an input, which is sampled from $z$\\ 
\hline

$KST$ & KS Test, which quantifies the distance between two distributions based on their empirical cumulative distribution functions (ECDF) and determines whether the incoming user traffic follows the observed data distribution or not\\ \hline
\textit{$P_{value}$} & Value determined by KS Test for each window of length $WS$ by comparing both generated and incoming user data\\ 
\hline
\textit{$D$} &  Discriminator built with multi-layer perceptron neural network to differentiate between the observed and generated data\\ \hline

\textit{$G$} & Generator built with the LSTM architecture to generate data close to the observed data\\ 
\hline

\textit{$P_{GAN-kstest}$} & Value determined by Kolmogorov-Smirnov Test (KS Test) between the generated and observed data during the training of GAN\\ 
\hline

\textit{$D_{score}$} & Range of discriminator score (i.e., between 0 and 1) obtained for the observed data during training of GAN\\ 
\hline

\textit{$Y_{predict}$} & Stores the $D$ prediction output in a list\\ 
\hline

\textit {$gg_{\text {data }}$} & Generated (or new) data from Generator $G$\\ 
\hline

\textit{$G({z})$} & Generator $G$, taking a vector $z_i$ of size $WS$ as an input, which is sampled from $z$ \\ 
\hline

\hline
\end{tabular}
\end{table*}

\begin{algorithm}
\caption{Predicting \ac{AI/ML} Model Retraining Using \acp{VAE}.}
\label{alg_VAE}
\begin{algorithmic}[1]
\State{\textbf{Input:} $ds$, $\mathcal{DE}$, 
$P_{VAE-kstest}$, $WS$, $z$}
\State{\textbf{Output:} Retrain the AI/ML model or not}
\While{data\_available}
\For {$i \gets 0$ to $\lfloor \frac{length({ds})}{WS} \rfloor$}
\State{$gd_{data} \gets \mathcal{DE}(z)$}
\State{$P_{value} \gets KST[gd_{data}, ds[i*WS\ to\ ((i+1)*WS-1)]]$}
\If{$P_{value}\ <\ P_{VAE-kstest}$}
\State{$Retrain\ the\ AI/ML\ model$}
\EndIf
\EndFor
\EndWhile
\end{algorithmic}
\end{algorithm}

Algorithm~\ref{alg_VAE} takes $ds$, $\mathcal{DE}$, $P_{VAE-kstest}$, $WS$, and a $z$ as inputs and predicts whether to retrain an \ac{AI/ML} model or not as an output using \acp{VAE}.
The incoming data stream $ds$ divides into consecutive chunks of data with a length of $WS$, and the \emph{decoder} ($\mathcal{DE}$) takes the latent space $z$ as input and generates data (i.e., $gd_{data}$) of length $WS$ (see lines 4-5). 
The $gd_{data}$ is compared with the data chunks $WS$ from the incoming user traffic using the \ac{KS Test} to determine changes in the incoming user traffic. 
The \ac{KS Test} computes the maximum distance between the \acp{CDF} of the two distributions (i.e., $gd_{data}$ and $ds$).
Here, \ac{KS Test} calculates a $P_{value}$, representing the probability that the $ds$ and the $gd_{data}$ are from the observed data. 
A lower $P_{value}$ indicates the arrival of significant changes in incoming user traffic. 
Whenever the $P_{value}$ is lower than $P_{VAE-kstest}$, the proposed approach triggers retraining of the \ac{AI/ML} model, indicating that the traffic from the incoming user is different from the underlying observed data (see lines 7-10).

\begin{algorithm}
\caption{Predicting AI/ML Model Retraining Using GANs.}
\label{alg_GAN}
\begin{algorithmic}[1]
\State{\textbf{Input:} $ds$,  $D$, $G$, $P_{GAN-kstest}$, $\mathcal{D}_{score}$, $WS$, $Noise$}
\State{\textbf{Output:} Retrain the AI/ML model or not}
\While {data\_available}
\For {$i \gets 0$ to $\lfloor \frac{length({ds})}{WS} \rfloor$}
\State {$Y_{predict} \gets D(ds[i*WS\ to\ ((i+1)*WS-1)])$} 
\For {$j \gets 0\ to\ WS$}
\If{$Y_{predict}[j]\ \textit{is\ in}\ \mathcal{D}_{score}$}
\State{$Do\_Nothing$}
\Else
\State{$Warning\ Zone$}
\EndIf
\State{$z \gets Noise$}
\State{$gg_{data} \gets G(z)$}
\State{$P_{value} \gets KST[gg_{data}, ds[i*WS\ to\ ((i+1)*WS-1)]]$}
\If{$P_{value}\ <\ P_{GAN-kstest}$}
\State{$Retrain\ the\ AI/ML\ model$}
\EndIf
\EndFor
\EndFor
\EndWhile
\end{algorithmic}
\end{algorithm}

Algorithm~\ref{alg_GAN} takes $ds$, $D$, $G$, $P_{GAN-kstest}$, $\mathcal{D}_{score}$, $WS$, and a $Noise$ as inputs and predicts whether to retrain an \ac{AI/ML} model or not as an output using \acp{GAN}. 
The Discriminator $D$ predicts whether the incoming user traffic belongs to the observed or not over each data sample in the data chunk (see lines 4-6). 
The $D$ assigns a score between $0$ and $1$ for each data sample, and if the score lies in the range of $\mathcal{D}_{score}$, no action will be taken (see lines 7-8). 
Otherwise, a $warning zone$ notifies by indicating changes in the incoming user traffic (see lines 9-10), and helps to collect data for model retraining, if required. 
The generator ($G$) takes the $Noise$ as input and generates data (i.e., $gg_{data}$) of length $WS$ (see lines 12-13). 
The $gg_{data}$ is compared with the data chunks from the incoming user traffic using the \ac{KS Test} to determine changes in the user traffic. 
The \ac{KS Test} is applied on the two distributions (i.e., $gg_{data}$ and $ds$) and calculates a $P_{value}$. 
Whenever the $P_{value}$ is lower than the $P_{GAN-kstest}$, the proposed approach triggers \ac{AI/ML} model retraining, indicating that the incoming user traffic is different from the underlying observed data (see lines 14-17). 
During the retraining, the previously trained \ac{AI/ML} model weights are updated based on the newly available data.

\begin{figure*}[h]
\centering
\includegraphics[width=1.0\linewidth]{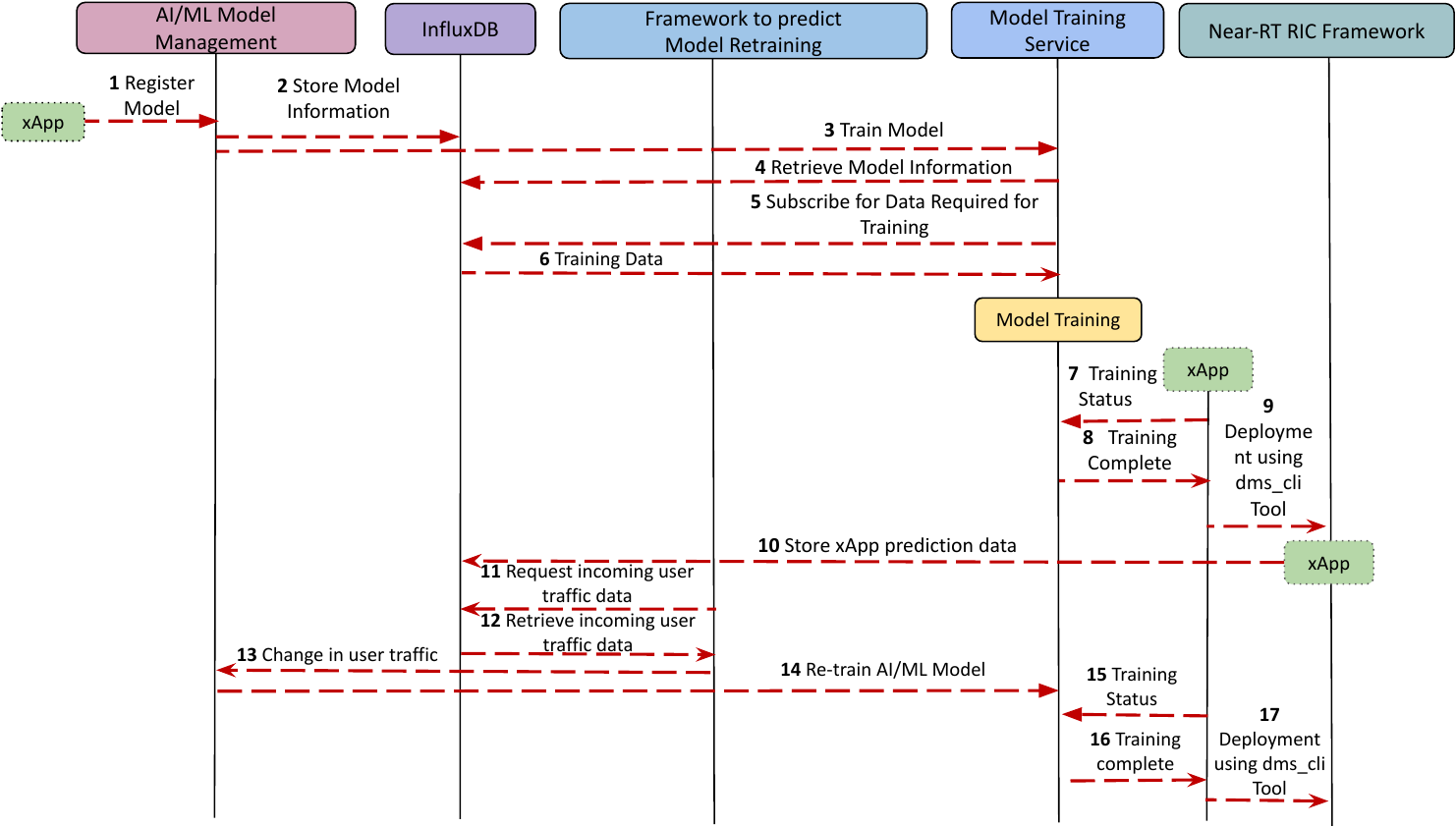}
\caption{Proposed Approach Workflow.}
\label{fig:arch_aiml_flow}
\end{figure*}

Figure~\ref{fig:arch_aiml_flow} depicts the workflow between the \ac{O-RAN} modules and the proposed predictive approach, detailed as follows: The \ac{AI/ML} model management registers model and stores its information into the data repository (1-2); the \ac{AI/ML} training service module initiates model training by retrieving requested data from the data repository (3-6); once the model training is complete, it will be deployed as xApp using $dms\_cli$ tool (7-9); the predicted user traffic through xApp can be stored into the data repository (10); the proposed predictive approach continuously retrieves the incoming user traffic data from the data repository and reports to \ac{AI/ML} model management through \ac{AI/ML} continues operation upon determining the changes in the user traffic (11-13); the \ac{AI/ML} model management triggers retraining based on the proposed predictive approach and updated model is deployed using $dms\_cli$ tool (15-17).


\section{Evaluation Setup}\label{sec:evalSetup}

The proposed predictive approach evaluated for two use cases: \emph{(i) Quality of Service (QoS)prediction~\cite{kousaridas2021qos}} over the \ac{OSC} platform,  and \emph{(ii) Network Slicing (NS)} using a real-time dataset~\cite{bonati2021intelligence}. 
Description of each use case and its corresponding experimental setup are as follows:

\begin{figure*}[h]
\centering
\includegraphics[width=1.0\linewidth]{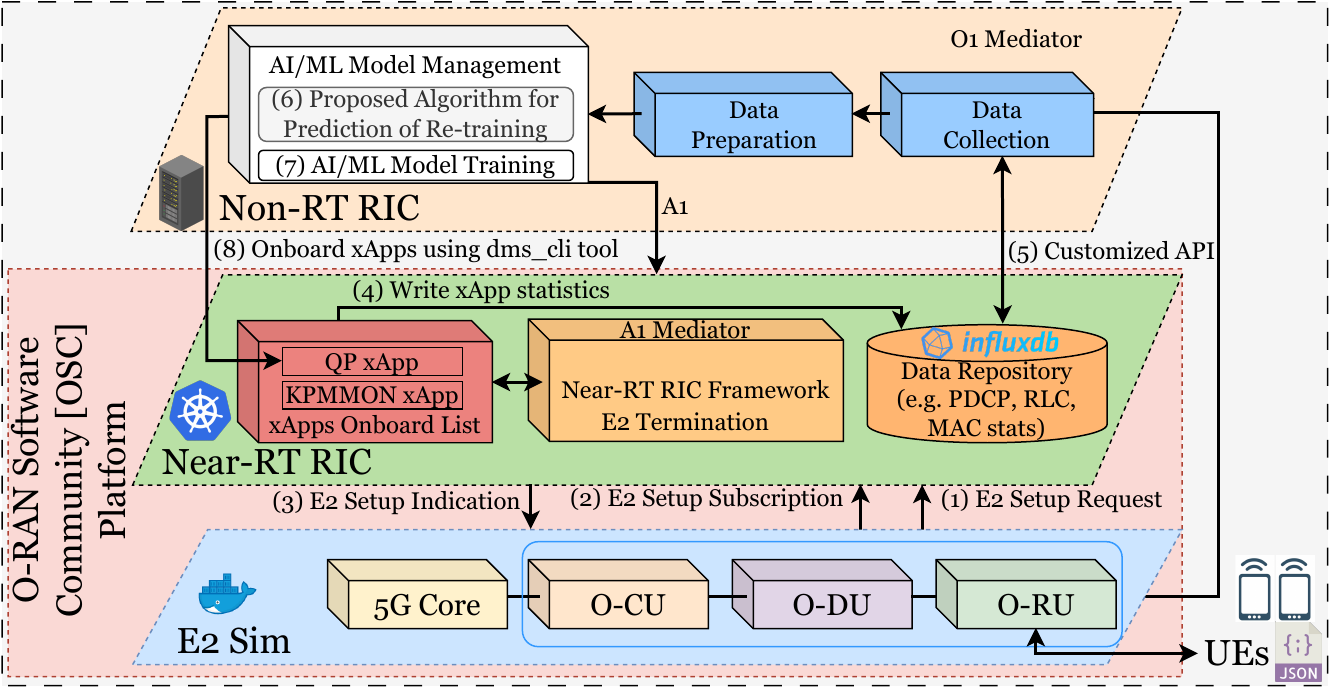}
\caption{\ac{QoS} experimental setup.}
\label{fig:exp_setup_eu}
\end{figure*}

\subsection{\ac{QoS} Use case}

The proposed approach is evaluated for \emph{\ac{QoS} prediction} use case, which focuses on predicting the service quality (i.e., throughput) provided by network operators to their customers~\cite{release18,ivanovic2012exploring}. Figure~\ref{fig:exp_setup_eu} depicts the experimental setup used to evaluate the \ac{QoS} prediction use case, in which the proposed approach is deployed at the \ac{Non-RT RIC} platform and integrated with the \ac{OSC} \ac{RIC} platform.

The \ac{OSC} \ac{Near-RT RIC} framework (F-release), an E2 simulator, and various open interfaces~\cite{OSC} are utilized within a \ac{K8s} pod. 
Other \ac{OSC} components like E2 manager, routing manager, subscription manager, app manager, and shared database (InfluxDB) are deployed as \ac{K8s} microservices. 
The \ac{Near-RT RIC} employs a \ac{KPMMON} xApp for monitoring \ac{RAN} layer statistics, and a \ac{QP} xApp using a three-layer \ac{LSTM}~\cite{eucnc} model for predicting \ac{QoS} values.

An \ac{OSC} E2 simulator consists of a 5G core, \ac{O-CU}, \ac{O-DU}, and \ac{O-RU}. 
The \ac{RAN} components establish communication with the \ac{Near-RT RIC} through the E2 control messages such as (1) E2 setup request; (2) E2 setup subscription; and (3) E2 setup indication as shown in Figure~\ref{fig:exp_setup_eu}. 
In addition, the E2 simulator can be connected with multiple \acp{UE} through a customized \ac{json} file. 

Note that the current F-release of \ac{OSC} does not include a complete implementation of the \ac{SMO} and the \ac{Non-RT RIC} frameworks. 
Therefore, to evaluate the \ac{QoS} prediction use case using the proposed approach, the missing functionalities in the \ac{OSC} platform are implemented and deployed on a bare metal server. 
For example, the $dms\_cli$ tool is used to onboard the trained model at the \ac{Near-RT RIC} as a microservice. 
Moreover, the proposed predictive approach is implemented as a \ac{REST} \ac{API} to monitor changes in user traffic by accessing real-time data (i.e., throughput) from the InfluxDB database through a customized \ac{API}. 

The \ac{Near-RT RIC} and E2 simulator are connected and integrated with the proposed predictive approach --- which used a \ac{Gen-AI} techniques to determine changes in the incoming user traffic and trigger retraining when necessary --- deployed in \ac{Non-RT RIC} to provide \ac{AI/ML} model management functionalities.

\begin{figure*}[h]
\centering
\includegraphics[width=1.0\linewidth]{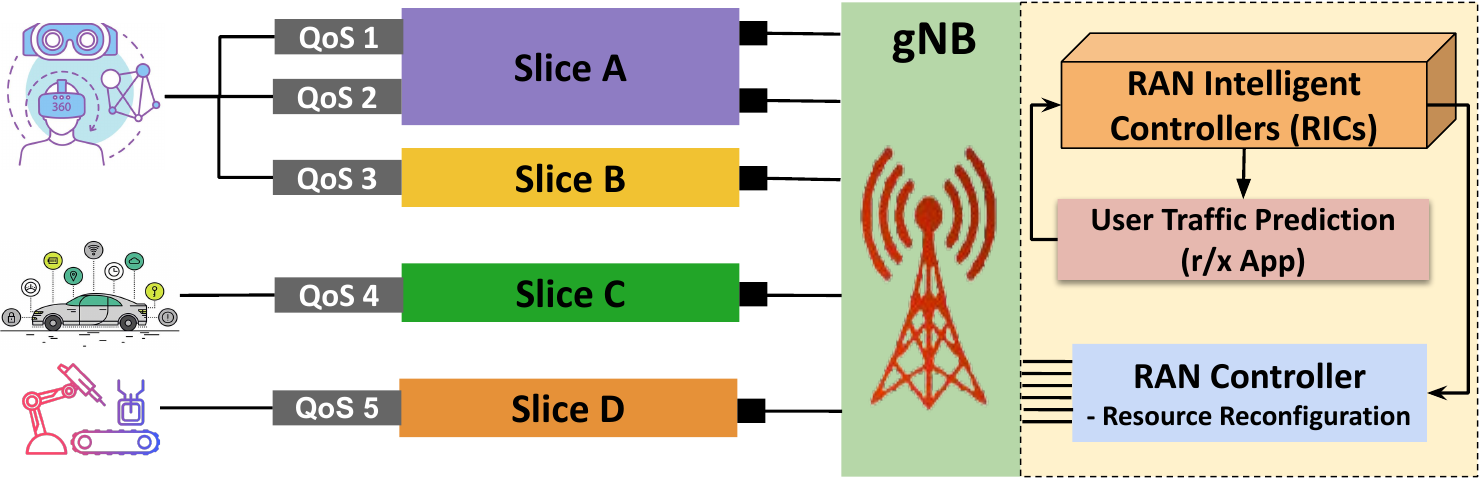}
\caption{Slicing use case~\cite{slicem}}
\label{fig:slice_usecase}
\end{figure*}

\subsection{Network Slicing Use case}

\ac{NS} is a key feature of \ac{B5G} networks, involves segmenting a single physical network into virtual slices, each tailored to meet the unique demands of specific use cases. 
The ability to allocate resources, configure parameters, and provide distinct service characteristics for each slice offers flexibility and efficiency. 
Moreover, enabling \ac{NS} at the \ac{RAN} --- an interface between \acp{UE} and the core network--- is important for optimizing resource utilization and ensuring a seamless end-to-end slicing architecture~\cite{bonati2021intelligence}.

\ac{NS}, can vary based on the user traffic demands as the \ac{B5G} networks continue to evolve and \ac{MNO} has to adapt to the user demands effectively~\cite{slicem, ericssonslicing, raftopoulos2024drlbased}. 
\acp{MNO} can employ \ac{AI/ML} models to predict the incoming user traffic through \acp{RIC} in the form of r/xApps and allocate the resources using \ac{RAN} controller messages as shown in Figure~\ref{fig:slice_usecase}. 
However, the employed \ac{AI/ML} models may not perform well as network service demands continue to evolve by introducing a new \ac{NS} traffic.
The proposed approach determines changes in the user traffic and predicts the need for model retraining to support the evolving \ac{NS} demands.

To evaluate the \ac{NS} use case with the proposed predictive approach considered a real-time dataset collected from the Colosseum testbed ~\cite{bonati2021intelligence}. 
The dataset is collected from the urban scenario of Rome, Italy, with 4 \acp{gNB} and 40 \acp{UE}.
The locations of the \acp{gNB} cover an area of 0.11 $km^2$ (i.e., square kilometer). 

\begin{table*}[bp]
    \centering
    \caption{\ac{GAN}, \ac{VAE}, and \ac{LSTM} model description}
    \label{tab:gan_vae_lstm_description}
    \begin{tabular}{|l|l|l|l|}
    \hline
    \multirow{8}{5em}{\ac{GAN}} & \multirow{4}{5em}{Generator} & Architecture & \ac{LSTM} layer + 3 \ac{FC} layers \\
    \cline{3-4} & & Output Activation & Sigmoid \\
    \cline{3-4} & & Loss Function & Binary Cross-Entropy \\
    \cline{3-4} & & Learning Rate & 0.001 \\
    \cline{2-4} & \multirow{4}{*}{Discriminator} & Architecture & 3 \ac{FC} layers \\
    \cline{3-4} & & Output Activation & Sigmoid \\
    \cline{3-4} & & Loss Function & Binary Cross-Entropy \\
    \cline{3-4} & & Learning Rate & 0.001 \\
    \hline
    \multirow{8}{5em}{\ac{VAE}} & \multirow{4}{*}{Encoder} & Architecture & \ac{MLP} \\
    \cline{3-4} & & Latent Dimension & 32 \\
    \cline{3-4} & & Loss Function & \ac{KL} Divergence \\
    \cline{3-4} & & Learning Rate & 0.001 \\
    \cline{2-4} & \multirow{4}{*}{Decoder} & Architecture & \ac{MLP} \\
    \cline{3-4} & & Activation function & Sigmoid \\
    \cline{3-4} & & Loss Function & \ac{MSE} \\
    \cline{3-4} & & Learning Rate & 0.001 \\
    \hline
    \multicolumn{2}{|l|}{\multirow{4}{*}{\ac{LSTM}}}  & Architecture & 3 \ac{LSTM} layers with 100 hidden units \\
    \cline{3-4}\multicolumn{2}{|l|}{} & Activation function & \ac{ReLU} \\
    \cline{3-4}\multicolumn{2}{|l|}{}  & Loss Function & \ac{MSE} \\
    \cline{3-4}\multicolumn{2}{|l|}{}  & Learning Rate & 0.001 \\
    \hline
    \end{tabular}
\end{table*}
The Colosseum dataset is based on a multi-slice scenario in which \acp{UE} are assigned to a slice of the network and request three different traffic types: high-capacity \ac{eMBB}, \ac{uRLLC}, and \ac{mMTC}. 
The \acp{gNB} serves each slice with dedicated scheduling policies. 
The scenario considered in this dataset concerns pedestrian user mobility with time-varying path loss and channel conditions. 
Traffic among \acp{gNB} and \acp{UE} is generated through the \ac{TGEN}, configured to send different traffic types to \acp{UE} of different slices such as: 
 $(i)$ \ac{eMBB}: 1 Mb/s constant bit rate traffic; 
 $(ii)$ \ac{uRLLC}: Poisson traffic, with 10 pkt/s of 125 bytes; and 
 $(iii)$ \ac{mMTC}: Poisson traffic, with 30 pkt/s of 125 bytes.

Three scenarios are available based on \acp{UE} mobility and the \acp{UE} distance from the \ac{gNB}~\cite{bonati2021intelligence} --- Static Close, Static Far, and Slow Close.

The available data sets scenario details are as follows:
\begin{itemize}
    \item Slow Close --- The \acp{UE} are uniformly distributed within $20\ m$ of each \ac{gNB} while moving $3 m/s$.
     \item Static Close --- The \acp{UE} are uniformly distributed within $20\ m$ of each \ac{gNB} with no mobility.
    \item Static Far --- The \acp{UE} are uniformly distributed within $100\ m$ of each \ac{gNB} with no mobility.
\end{itemize}

The performance results are obtained for all the above three scenarios and understood the proposed predictive approach is able to perform equally independent of the UE distance and the mobility. Due to the proposed predictive approach adaptability ---  \acp{GAN} and \acp{VAE} are capable enough to learn the complex data distributions/patterns.

In this paper, Slow Close is considered for the performance valuation.

\subsection{Performance comparison}
The proposed predictive approach is compared with the other two approaches to evaluate the performance. 
The other two considered approaches are as follows:

\begin{itemize}
   \item \textbf{Threshold Approach~\cite{mekrachecombining}:} 
    Triggers the model retraining whenever the considered \ac{AI/ML} model performance metric exceeds the pre-defined threshold. The \ac{RMSE} over each data chunk of size $WS$ serves as the performance metric for the threshold approach.
   
    \item \textbf{Predictive Approach
    [\ac{LOF}]~\cite{Gudepu2023AdaptiveRO}:} 
    Predicts when to retrain the AI/ML model by exploiting a \ac{LOF} classifier that calculates the local density deviation for each incoming user data sample and determines whether it belongs to the observed or new data. 
    If new user data arrive over a certain number of windows of size $WS$ --- which can be calculated as a function of the time taken to transmit the considered data samples from the data stream (i.e., $T_{ds}$) and the end-to-end delay of the application under consideration (i.e., $T_{e2e}$) --- then the predictive approach [\ac{LOF}] triggers retraining.

\end{itemize}

The performance of the proposed and considered approaches are evaluated with the following key metrics:

$(i)$ \textit{\ac{MRTT}:} It is defined as the time taken between the occurrence of a change in the user traffic and the \ac{AI/ML} model triggering for retraining. 
A shorter MRTT signifies the efficiency of an approach in promptly identifying changes in incoming user traffic. The \ac{MRTT} is crucial for assessing how quickly an approach can adapt to dynamic user behavior.

$(ii)$ \textit{\ac{MRPT}:} It is defined as the time taken to retrain an \ac{AI/ML} model and to replace with the existing model.
A shorter MRPT is essential for minimizing \ac{SLA} violations and addressing resource utilization issues arising due to the changes in the incoming user traffic.

\section{Results} \label{sec:results}

This section presents the results obtained for the use cases under consideration --- $(i)$ \ac{QoS} Prediction and $(ii)$ \ac{NS}. 
Additionally, a comparative analysis of the proposed approach with the other two approaches such as $(i)$ Predictive approach [\ac{LOF}] and $(ii)$ Threshold approach, are presented.
Also, the details of \ac{GAN}, \ac{VAE}, and \ac{LSTM} architecture are listed in Table~\ref{tab:gan_vae_lstm_description}.

\begin{figure*}[h]
 \centering
\includegraphics[width=0.99\linewidth]{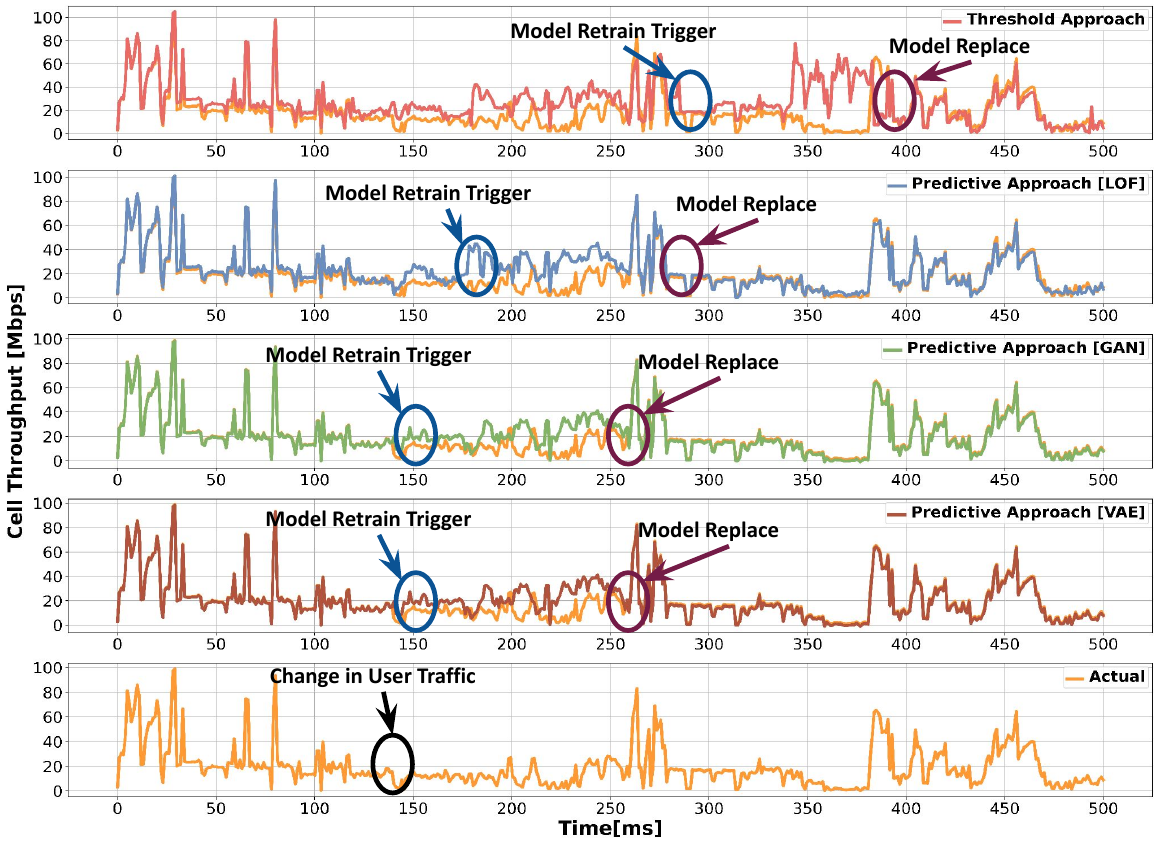}
\caption{Evaluation of QoS Prediction (QP) xApp.}
\label{fig:final_user_exp}
\end{figure*}

\subsection{QoS Prediction (QP) Use case}

The deployed \ac{QP} xApp is implemented using the \ac{LSTM} model and trained with data collected from \ac{KPMMON} xApp and other datasets~\cite{bonati2021intelligence, raca2020beyond}. Figure~\ref{fig:final_user_exp} depicts the performance of the considered approaches in predicting the throughput. 
To evaluate the \ac{QoS} use case, the input parameters for all the considered approaches are as listed in Table~\ref{tab:parameterlist_QoS}.
Here, the $P_{VAE-kstest}$ and $P_{GAN-kstest}$ values are determined during the training phase. Other parameters like $T_{ds}$ and $T_{e2e}$ are use case specific~\cite{5gverticals}, while $WS$ depends on operator requirement.
Here, the larger $WS$ value provides more stability by tolerating short-term changes in user traffic, whereas a smaller $WS$ value increases sensitivity to immediate changes in user traffic, allowing for a prompt response but potentially leading to frequent retraining.

\begin{table}[h]
    \centering
    \caption{Input parameters list of QoS use case}
    \label{tab:parameterlist_QoS}
    \begin{tabular}{|p{0.3\columnwidth}|p{0.3\columnwidth}|p{0.15\columnwidth}|}
        \hline
        \multirow{2}{*}{\textbf{Approach}} & \multirow{2}{*}{\textbf{Parameter}} & \multirow{2}{*}{\textbf{Value}} \\
        & & \\
        \hline
        \multirow{2}{0.2\columnwidth}{Threshold Approach} & \ac{RMSE} & 15 \\
        &  & \\
        \hline
        \multirow{3}{0.2\columnwidth}{Predictive Approach [\ac{LOF}]} & $T_{ds}$ & $20\ ms$ \\
        & $T_{e2e}$ &  $5\ ms$ \\ 
        & & \\
        \hline
        \multirow{3}{0.2\columnwidth}{Proposed Predictive Approach}& $WS$ & 10 \\
        & $P_{VAE-kstest}$ & 0.00182 \\
        & $P_{GAN-kstest}$ & 0.00206 \\
        \hline
    \end{tabular}
\end{table}

\begin{figure}[h]
\centering
\includegraphics[width=1.0\linewidth]{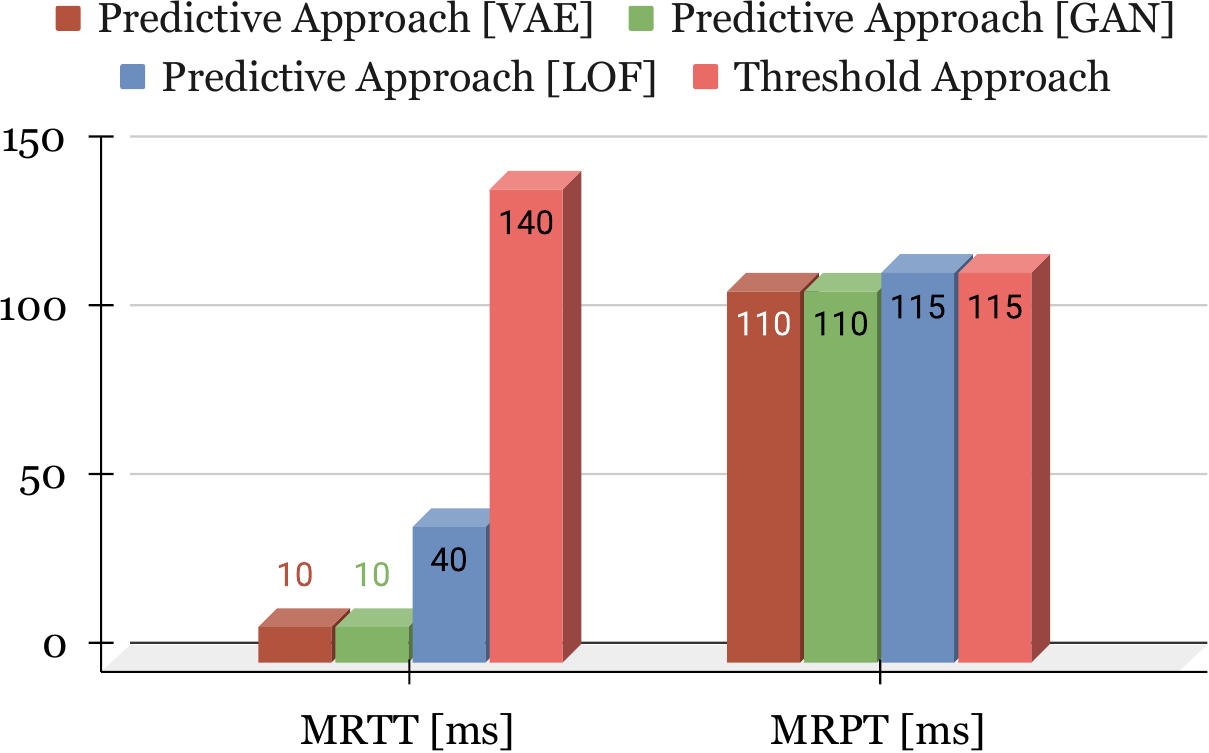}
\caption{Performance comparison of the considered approaches.}
\label{fig:proposed_approach_comparison}
\end{figure}

Figure~\ref{fig:final_user_exp} shows the evaluation of QP xApp as a function of cell throughput. Initially, from $0\ ms$ to $139\ ms$ ---- pedestrian traffic is considered, and $140\ ms$ onwards multiple \acp{UE} with different data rates are configured to create changes in the incoming user traffic (i.e., the bottom plot of Figure~\ref{fig:final_user_exp}). 
In this scenario, the \emph{threshold approach} identified changes in user traffic $270\ ms$ - $280\ ms$ and initiated retraining at $280\ ms$ upon the \ac{RMSE} exceeded the pre-defined threshold. The retrained (e.g., \ac{LSTM}) model is replaced at $395\ ms$ for continuous throughput prediction. 
Whereas, the predictive approach [\ac{LOF}] triggered retraining at $180\ ms$ after detecting changes in user traffic $140\ ms$ - $180\ ms$ and replaced at $295\ ms$. 
The proposed predictive approach --- both \ac{GAN} and \ac{VAE} trigger retraining at $150\ ms$ and replaced with a retrained model at $260\ ms$ and $260\ ms$ to predict the throughput values accurately. 

\begin{figure}[h]
\centering
\includegraphics[width=\linewidth]{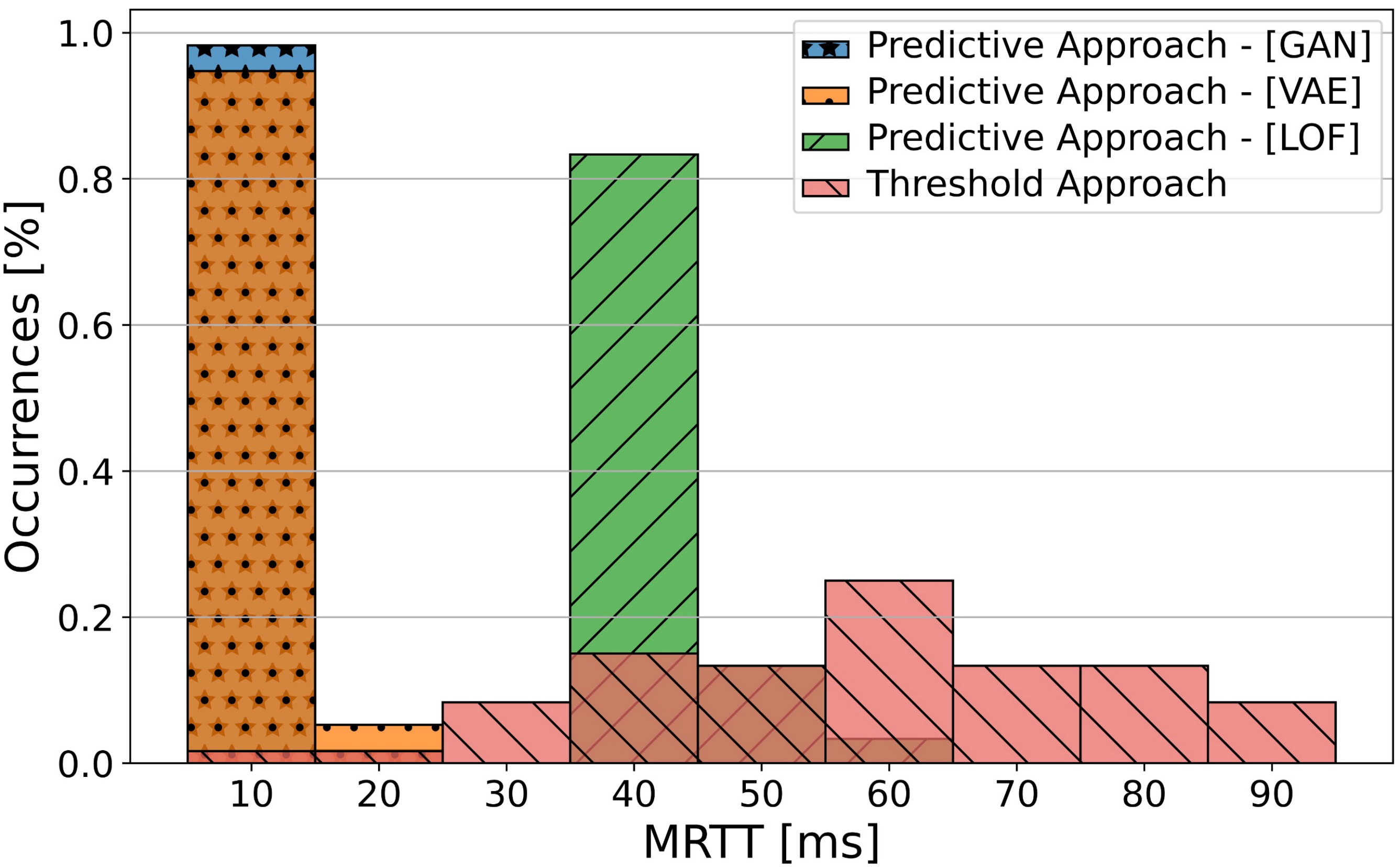}
\caption{Sampled PMF of model retraining trigger time.}
\label{fig:pmf_plot}
\end{figure}

Figure~\ref{fig:proposed_approach_comparison} presents both \ac{MRTT} and \ac{MRPT} for each of the considered approaches. The shorter \ac{MRTT} shows the proposed approach efficiently detects user traffic changes and helps decrease the retraining time due to fewer samples considered for retraining. However, this cannot be identified from the captured results (i.e., \ac{MRPT}) due to the lower number of samples in all the considered scenarios. Moreover, this could differentiate with larger samples required to be retrained.

Figure~\ref{fig:pmf_plot} shows a sampled \ac{PMF} between the time required to identify the occurrence of a change in user traffic (i.e., \ac{MRTT}) and the frequency of such occurrences. On the Y-axis of occurrences represent the percentage of instances out of $60$ (1-hour duration) to trigger retraining after the occurrence of an actual traffic change. 
Whereas, the X-axis is a discrete variable as the \ac{MRTT} will be a factor of $WS$. The threshold approach exhibits a maximum of $25\%$ occurrences at $60\ ms$, while the predictive approach [\ac{LOF}] shows approximately $85\%$ occurrences at $40\ ms$. 
In contrast, the proposed predictive approach [\ac{GAN}] and [\ac{VAE}] achieves approximately $98\%$ occurrences at $10\ ms$, and $95\%$ occurrences at $10\ ms$, respectively.
Thus, the proposed predictive approach --- \ac{GAN} and \ac{VAE} --- effective determination of user traffic changes.


\subsection{Network Slicing (\ac{NS}) Use case}

\begin{figure*}[h]
 \centering
\includegraphics[width=1.0\linewidth]{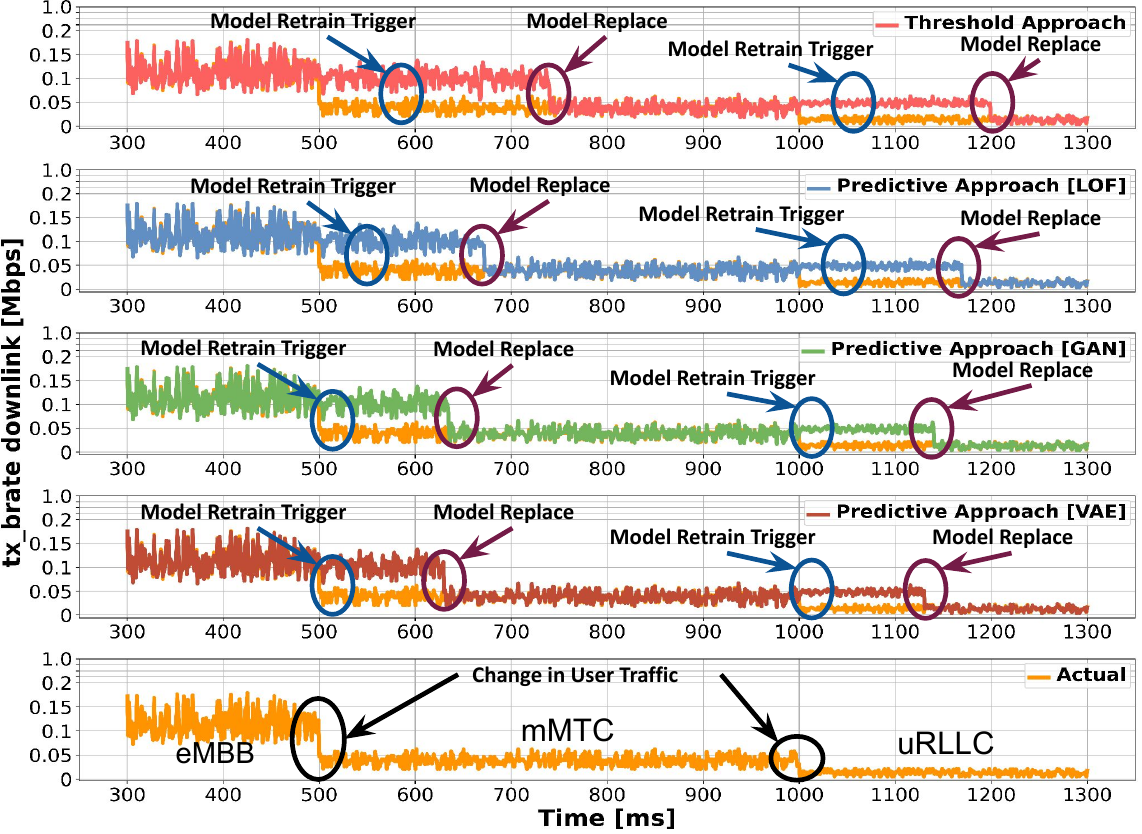}
\caption{Slow\_Close - \ac{eMBB} $ \longrightarrow $ \ac{mMTC} $ \longrightarrow $ \ac{uRLLC}.}
\label{fig:Slow_Close_LSTM}
\end{figure*}

Figure~\ref{fig:Slow_Close_LSTM} shows the considered Slow Close with the considered approaches in predicting downlink $tx\_brate$. The input parameters for all the considered approaches are listed in Table~\ref{tab:parameterlist_NS}.

\begin{table}[h]
    \centering
    \caption{Input parameters list for \ac{NS} use case}
    \label{tab:parameterlist_NS}
    \begin{tabular}{|p{0.2\columnwidth}|p{0.5\columnwidth}|p{0.15\columnwidth}|}
        \hline
        \multirow{2}{*}{\textbf{Approach}} & \multirow{2}{*}{\textbf{Parameter}} & \multirow{2}{*}{\textbf{Value}} \\
        & & \\
        \hline
        \multirow{2}{0.2\columnwidth}{Threshold Approach} & \ac{RMSE} (\ac{eMBB} $ \longrightarrow $ \ac{mMTC}) & 15 \\
        & \ac{RMSE} (\ac{mMTC} $ \longrightarrow $ \ac{uRLLC}) & 5\\
        & \ac{RMSE} (\ac{uRLLC} $ \longrightarrow $ \ac{eMBB}) & 15 \\
        \hline
        \multirow{3}{0.2\columnwidth}{Predictive Approach [\ac{LOF}]} & $T_{ds}$ & $20\ ms$ \\
        & $T_{e2e}$ &  $5\ ms$ \\ 
        & & \\
        \hline
        \multirow{3}{0.2\columnwidth}{Proposed Predictive Approach}& $WS$ & 10 \\
        & $P_{VAE-kstest}$ (Slow Close) & 0.01235 \\
        & $P_{GAN-kstest}$ (Slow Close) & 0.01568 \\
        \hline
    \end{tabular}
\end{table}

As shown in Figure~\ref{fig:Slow_Close_LSTM}, initially, the \ac{eMBB} slice traffic is considered and user traffic change is introduced at $500\ ms$ by streaming the \ac{mMTC} slice traffic from $500\ ms$ onwards. 
Here, the threshold approach triggers retraining at $590\ ms$ upon detecting changes $500\ ms$ - $590\ ms$ with model replacement at $750\ ms$. 
Similarly, the predictive approach [\ac{LOF}] initiates retraining at $540\ ms$ after detecting changes $500\ ms$ - $540\ ms$ with model replacement at $680\ ms$. 
In contrast, the proposed predictive approach using both \ac{GAN} and \ac{VAE} triggers retraining at $510\ ms$ after identifying changes in user traffic, with model replacements at $628\ ms$ and $630\ ms$, respectively.

Furthermore, \acp{UE} switches from the \ac{mMTC} slice to the \ac{uRLLC} slice, observable from $1000\ ms$. 
The threshold approach detects traffic changes $1000\ ms$ - $1060\ ms$, triggers retraining at $1060\ ms$ when the \ac{RMSE} surpasses the threshold and the model replacement at $1200\ ms$.
Also, the predictive approach [\ac{LOF}] triggers retraining at $1040\ ms$ after detecting changes $1000\ ms$ - $1040\ ms$, with model replacement at $1180\ ms$.
Whereas, the proposed predictive approach using both \ac{GAN} and \ac{VAE} starts retraining at $1010\ ms$ upon identifying changes in user traffic, with model replacements at $1125\ ms$ and $1130\ ms$, respectively, to ensure accurate throughput predictions.

\begin{figure}[h]
\centering
\includegraphics[width=1.0\linewidth]{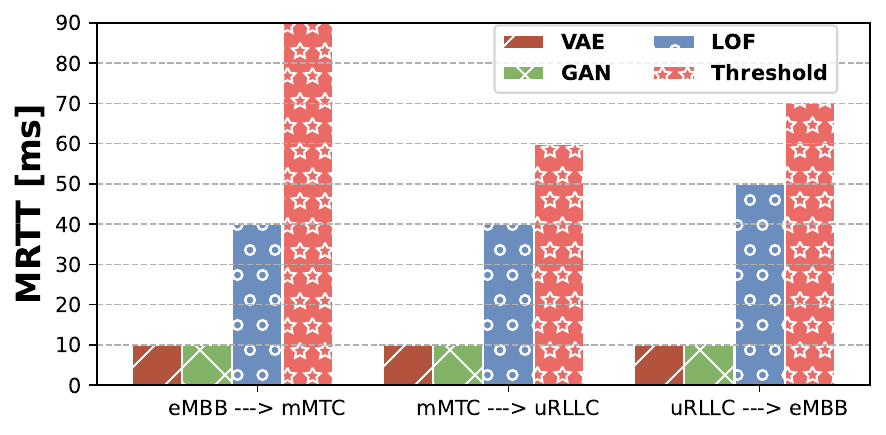}
\caption{Slow close}
\label{fig:Slow_Close}
\end{figure}

Figure~\ref{fig:Slow_Close} illustrates the \ac{MRTT} for each of the approaches for the Slow Close scenario based on Figure~\ref{fig:Slow_Close_LSTM}. It can be observed that the \ac{MRTT} for the proposed predictive approach based on \ac{Gen-AI} techniques --- \ac{GAN} and \ac{VAE} is smaller compared to the other approaches, due to the following reasons: $(i)$ \ac{Gen-AI} techniques are capable enough in adapting to dynamic network conditions, traffic patterns, and user behaviour;
$(ii)$ The proposed predictive approach employed the \ac{KS Test} to determine distribution similarity, which enables the promptly determining user traffic changes. 

In the case of the threshold approach, the \ac{RMSE} values are low during the initial stage of incoming new user traffic, however, in the later stages the \ac{RMSE} values exceed the pre-defined thresholds. Thus, the threshold approach always exhibits the higher \ac{MRTT} when compared to other approaches. As stated earlier, the predictive approach [\ac{LOF}] operates mainly based on the selected number of windows with size $WS$, hence the approach must wait for the considered number of windows before triggering the retraining, which leads to the higher \ac{MRTT} when compared to proposed predictive approaches.

In summary, the proposed predictive approach based on the \ac{Gen-AI} outperforms the other approaches for use cases such as \ac{QoS} prediction and \ac{NS}.
The choice of these use cases is driven by their vital role in \ac{B5G} networks. 
\ac{QoS} prediction ensures consistent network performance, while \ac{NS} is essential to meet the user demands without violating \acp{SLA}. 
Both use cases demonstrate the adaptability and effectiveness of the proposed predictive approach.
A performance comparison reveals that the changes in user traffic are effectively adapted by the proposed predictive approach --- \ac{Gen-AI} compared to the predictive approach [\ac{LOF}] and the threshold approach.

\section{Complexities and Challenges}
\label{sec:challenges}
This section explains the complexities and challenges of dealing with \ac{Gen-AI}.

\subsection{Model Complexity}

Model complexity involves the number of parameters, features, or rules within \ac{Gen-AI} techniques to learn from data and make predictions accurately. 
The complexity of \ac{Gen-AI} introduces potential difficulties during implementation and integration into the \ac{RAN} through \acp{RIC}. 
Effectively managing \ac{Gen-AI} complexity requires specialized expertise in deployment and maintenance. Overcoming the challenge of model complexity is essential for ensuring the successful and robust integration of \ac{Gen-AI} in \ac{B5G} networks.

\subsection{Model Generalization}

Model generalization refers to the capability of a \ac{Gen-AI} technique to produce meaningful and accurate outputs not only for the trained data but also for new/unseen data --- drawn from a similar distribution. 
Model generalization emerges as a critical concern in the realm of \ac{Gen-AI} implementation for \ac{B5G} networks --- when the model has limited generalization, the model may lead to incorrect predictions for new/unseen data. Thus, ensuring these \ac{Gen-AI} techniques can effectively generalize their learned patterns and behaviors for diverse scenarios/use cases. 
In \ac{B5G} networks, \ac{Gen-AI} models must exhibit a robust ability to adapt and perform consistently well under various conditions. 
Addressing this challenge is imperative for unlocking the full potential of \ac{Gen-AI} in contributing to the adaptability and efficiency of \ac{B5G} networks.

\subsection{Model Interpretation}

Model interpretation involves the process of understanding and explaining the outputs and behavior of \ac{Gen-AI} techniques.
Wireless systems need to be engineered with justifiable results and performance guarantees when integrated with different network components and requirements.
Unfortunately, most of the existing \ac{Gen-AI} techniques follow a black-box approach without explaining why and how the approach led to the given outcome.
Developing \ac{XAI} with interpretable and predictable outcomes is one of the key challenges for applying \ac{AI} to \ac{B5G} networks.

\subsection{Computational Intensity}

The proposed predictive approach --- \ac{Gen-AI} outperforms the other approaches, however, note that \ac{Gen-AI} requires a higher computational cost compared to other approaches. 
Mitigating the computational overhead is crucial for the adoption of \ac{Gen-AI} models, particularly when considering deployment at the network edge. 
Exploring strategies to compress \ac{Gen-AI} techniques or efficiently distribute the computation becomes imperative to balance performance excellence and resource efficiency in real-time applications.

\section{Conclusions}\label{sec:conclusions}

Beyond 5G (B5G) networks are expected to bring transformations in the next-generation networks by utilizing Artificial Intelligence/Machine Learning (AI/ML) techniques.
However, maintaining performance consistency with the AI/ML model is crucial for different use cases such as massive Machine Type Communications (mMTC), ultra-Reliable Low Latency Communications (uRLLC), and enhanced Mobile Broadband (eMBB).  
In this paper, the proposed predictive approach utilizing \ac{Gen-AI} --- \ac{VAE} and \ac{GAN} --- to predict the \ac{AI/ML} model retraining in order to maintain \ac{AI/ML} model consistency.
The proposed predictive approach is evaluated by implementing on the \ac{OSC} platform and a real-time dataset for two different use cases --- \emph{QoS Prediction} and \emph{Network Slicing}.
The performance results showed that the proposed predictive approach outperforms the state-of-the-art --- predictive approach [\ac{LOF}] and threshold approach --- by detecting user traffic changes promptly after they occurr, thus helps in reducing \ac{SLA} violations significantly.
In particular, control loop time (i.e., model retraining trigger time, model retrain and replacement time) reduction in terms of microseconds can be achieved. 
Based on this study, the potential future directions of this work are as follows: 
$(i)$ \emph{Exploring the \ac{RL}}:
Formulating \ac{AI/ML} model retraining as a decision problem and investigating \ac{RL} algorithms for efficient decision-making; and
$(ii)$ \emph{Incorporating \ac{XAI}}:
Integrating \ac{XAI} to enhance model interpretability, providing insights into the reasons behind \ac{Gen-AI} predictions.

\bibliographystyle{elsarticle-num}
\bibliography{myrefs}

\begin{thebibliography}{10}
\expandafter\ifx\csname url\endcsname\relax
  \def\url#1{\texttt{#1}}\fi
\expandafter\ifx\csname urlprefix\endcsname\relax\def\urlprefix{URL }\fi
\expandafter\ifx\csname href\endcsname\relax
  \def\href#1#2{#2} \def\path#1{#1}\fi

\bibitem{bartsiokas2022ml}
I.~A. Bartsiokas, P.~K. Gkonis, D.~I. Kaklamani, I.~S. Venieris, {ML-Based Radio Resource Management in 5G and Beyond Networks: A Survey}, IEEE Access 10 (2022) 83507--83528.

\bibitem{khan2023ai}
N.~A. Khan, S.~Schmid, {AI-RAN in 6G Networks State-of-the-Art and Challenges}, IEEE Open Journal of the Communications Society (2023).

\bibitem{oranwg2}
{O-RAN Working Group 2}, {O-RAN AI/ML Workflow Description and Requirements - v1.01, Technical Specification, 2023}.

\bibitem{aryal2023open}
N.~Aryal, E.~Bertin, N.~Crespi, {Open Radio Access Network challenges for Next Generation Mobile Network}, in: 26th Conference on Innovation in Clouds, Internet and Networks and Workshops (ICIN), IEEE, 2023, pp. 90--94.

\bibitem{manias2023model}
D.~M. Manias, A.~Chouman, A.~Shami, {Model Drift in Dynamic Networks}, IEEE Communications Magazine (2023).

\bibitem{gudepu2023exploiting}
V.~Gudepu, V.~R. Chintapalli, L.~Valcarenghi, K.~Kondepu, {Exploiting Drift Detection Techniques for Next Generation Radio Access Networks}, in: 15th International Conference on COMmunication Systems \& NETworkS (COMSNETS), 2023, pp. 489--491.

\bibitem{slicem}
{Cisco}, {Quality of Service Configuration Guide, Cisco IOS XE 17.x}, accessed on 3 Feb 2024. [Online]. Available: \newline \url{https://www.cisco.com/c/en/us/td/docs/routers/ios\newline/config/17-x/qos/b-quality-of-service \newline/m\_qos-apply.html}.

\bibitem{5gverticals}
{TM forum}, {5G for Vertical Industries} (Aug, 2020).

\bibitem{al2020prediction}
A.~H.~A. Muktadir, V.~P. Kafle, {Prediction and Dynamic Adjustment of Resources for Latency-Sensitive Virtual Network Functions}, in: 23rd Conference on Innovation in Clouds, Internet and Networks and Workshops (ICIN), 2020, pp. 235--242.

\bibitem{mekrachecombining}
A.~Mekrache, K.~Boutiba, A.~Ksentini, {Combining Network Data Analytics Function and Machine Learning for Abnormal Traffic Detection in Beyond 5G}.

\bibitem{samdanis2023ai}
K.~Samdanis, A.~N. Abbou, J.~Song, T.~Taleb, {AI/ML Service Enablers \& Model Maintenance for Beyond 5G Networks}, IEEE Network (2023) 1--10.

\bibitem{Gudepu2023AdaptiveRO}
V.~Gudepu, V.~R. Chintapalli, P.~Castoldi, L.~Valcarenghi, B.~R. Tamma, K.~Kondepu, {Adaptive Retraining of AI/ML Model for Beyond 5G Networks: A Predictive Approach}, IEEE 9th International Conference on Network Softwarization (NetSoft) (2023) 282--286.

\bibitem{ganretrain}
V.~Gudepu, B.~Chirumalli, V.~R. Chintapalli, P.~Castoldi, L.~Valcarenghi, K.~Kondepu, {Generative Adversarial Networks-Based AI/ML Model Adaptive Retraining for Beyond 5G Networks}, in: IEEE 28th European Wireless (EW) conference, 2024.

\bibitem{karapantelakis2023generative}
A.~Karapantelakis, P.~Alizadeh, A.~Alabassi, K.~Dey, A.~Nikou, {Generative AI in mobile networks: a survey}, Annals of Telecommunications (2023) 1--19.

\bibitem{navidan2021generative}
H.~Navidan, P.~F. Moshiri, M.~Nabati, R.~Shahbazian, S.~A. Ghorashi, V.~Shah-Mansouri, D.~Windridge, {Generative Adversarial Networks (GANs) in networking: A comprehensive survey \& evaluation}, Computer Networks 194 (2021) 108149.

\bibitem{OSC}
{O-RAN software community (OSC)}.
\newblock \href{https://wiki.o-ran-sc.org/display/ORAN/\newline O-RAN+Software+Community}{[link]}.
\newline\urlprefix\url{https://wiki.o-ran-sc.org/display/ORAN/\newline O-RAN+Software+Community}

\bibitem{bonati2021intelligence}
L.~Bonati, S.~D'Oro, M.~Polese, S.~Basagni, T.~Melodia, {Intelligence and learning in O-RAN for data-driven NextG cellular networks}, IEEE Communications Magazine 59~(10) (2021) 21--27.

\bibitem{bank2023autoencoders}
D.~Bank, N.~Koenigstein, R.~Giryes, Autoencoders, Machine learning for data science handbook: data mining and knowledge discovery handbook (2023) 353--374.

\bibitem{raiber2017kullback}
F.~Raiber, O.~Kurland, {Kullback-Leibler Divergence Revisited}, in: Proceedings of the ACM SIGIR International Conference on Theory of Information Retrieval, 2017, pp. 117--124.

\bibitem{ayanoglu2022machine}
E.~Ayanoglu, K.~Davaslioglu, Y.~E. Sagduyu, {Machine Learning in NextG Networks via Generative Adversarial Networks}, IEEE Transactions on Cognitive Communications and Networking 8~(2) (2022) 480--501.

\bibitem{kaloxylos2021ai}
A.~Kaloxylos, A.~Gavras, D.~Camps, M.~Ghoraishi, H.~Hrasnica, {AI and ML--Enablers for beyond 5G Networks}, White Paper G (2021).

\bibitem{chintapalli2022wip}
V.~R. Chintapalli, V.~Gudepu, K.~Kondepu, A.~Sgambelluri, A.~Franklin, B.~R. Tamma, P.~Castoldi, L.~Valcarenghi, {WIP: Impact of AI/ML Model Adaptation on RAN Control Loop Response Time}, in: 2022 IEEE 23rd International Symposium on a World of Wireless, Mobile and Multimedia Networks (WoWMoM), IEEE, 2022, pp. 181--184.

\bibitem{marsaglia2003evaluating}
G.~Marsaglia, W.~W. Tsang, J.~Wang, {Evaluating Kolmogorov's distribution}, Journal of statistical software 8 (2003) 1--4.

\bibitem{kousaridas2021qos}
A.~Kousaridas, R.~P. Manjunath, J.~Perdomo, C.~Zhou, E.~Zielinski, S.~Schmitz, A.~Pfadler, {QoS Prediction for 5G Connected and Automated Driving}, IEEE Communications Magazine 59~(9) (2021) 58--64.

\bibitem{release18}
{Qualcomm}, {Setting off the 5G Advanced evolution} (2022).

\bibitem{ivanovic2012exploring}
D.~Ivanovi{\'c}, M.~Carro, M.~Hermenegildo, {Exploring the Impact of Inaccuracy and Imprecision of QoS Assumptions on Proactive Constraint-Based QoS Prediction for Service Orchestrations}, in: 2012 4th International Workshop on Principles of Engineering Service-Oriented Systems (PESOS), IEEE, 2012, pp. 29--35.

\bibitem{eucnc}
V.~Reddy~Chintapalli, K.~Kondepu, A.~Sgambelluri, A.~Franklin~A, B.~Reddy~Tamma, P.~Castoldi, L.~Valcarenghi, {Orchestrating Edge- and Cloud-based Predictive Analytics Services}, in: Proc. of EuCNC, 2020, pp. 214--218.

\bibitem{ericssonslicing}
{Ayodele Damola, Mathias Sintorn, Ajay Gautam}, {How to leverage intent-based automation with AI/ML for 5G RAN, ERICSSON BLOG, 2023}.

\bibitem{raftopoulos2024drlbased}
R.~Raftopoulos, S.~D'Oro, T.~Melodia, G.~Schembra, {DRL-based Latency-Aware Network Slicing in O-RAN with Time-Varying SLAs} (2024).
\newblock \href {http://arxiv.org/abs/2401.05042} {\path{arXiv:2401.05042}}.

\bibitem{raca2020beyond}
D.~Raca, D.~Leahy, C.~J. Sreenan, J.~J. Quinlan, {Beyond Throughput, The Next Generation: A 5G Dataset with Channel and Context Metrics}, in: Proceedings of the 11th ACM multimedia systems conference, 2020, pp. 303--308.

\end{thebibliography}

\begin{IEEEbiography}[{\includegraphics[width=1.05in,height=1.5in,keepaspectratio]{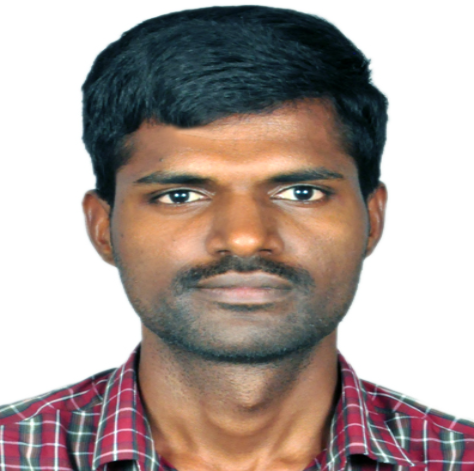}}]{VENKATESWARLU GUDEPU~} (Graduate Student Member, IEEE) received his B.Tech degree in computer science from Rajiv Gandhi University of Knowledge Technologies-Nuzvid (RGUKT-Nuzvid), India in 2017, and the M.Tech in Artificial Intelligence (AI) from National Institute of Technology (NIT-Uttarakhand), India. He is currently pursuing the Ph.D. in computer science and engineering with the Indian Institute of Technology - Dharwad (IIT Dharwad), India. His area of research interest includes 5G and Beyond technology, AI/ML for Networks, O-RAN standards implementation, and Energy sustainability for B5G Networks.
\end{IEEEbiography}

\vspace{-0.5cm}

\begin{IEEEbiography}[{\includegraphics[width=1.25in,height=1.05in,clip,keepaspectratio]{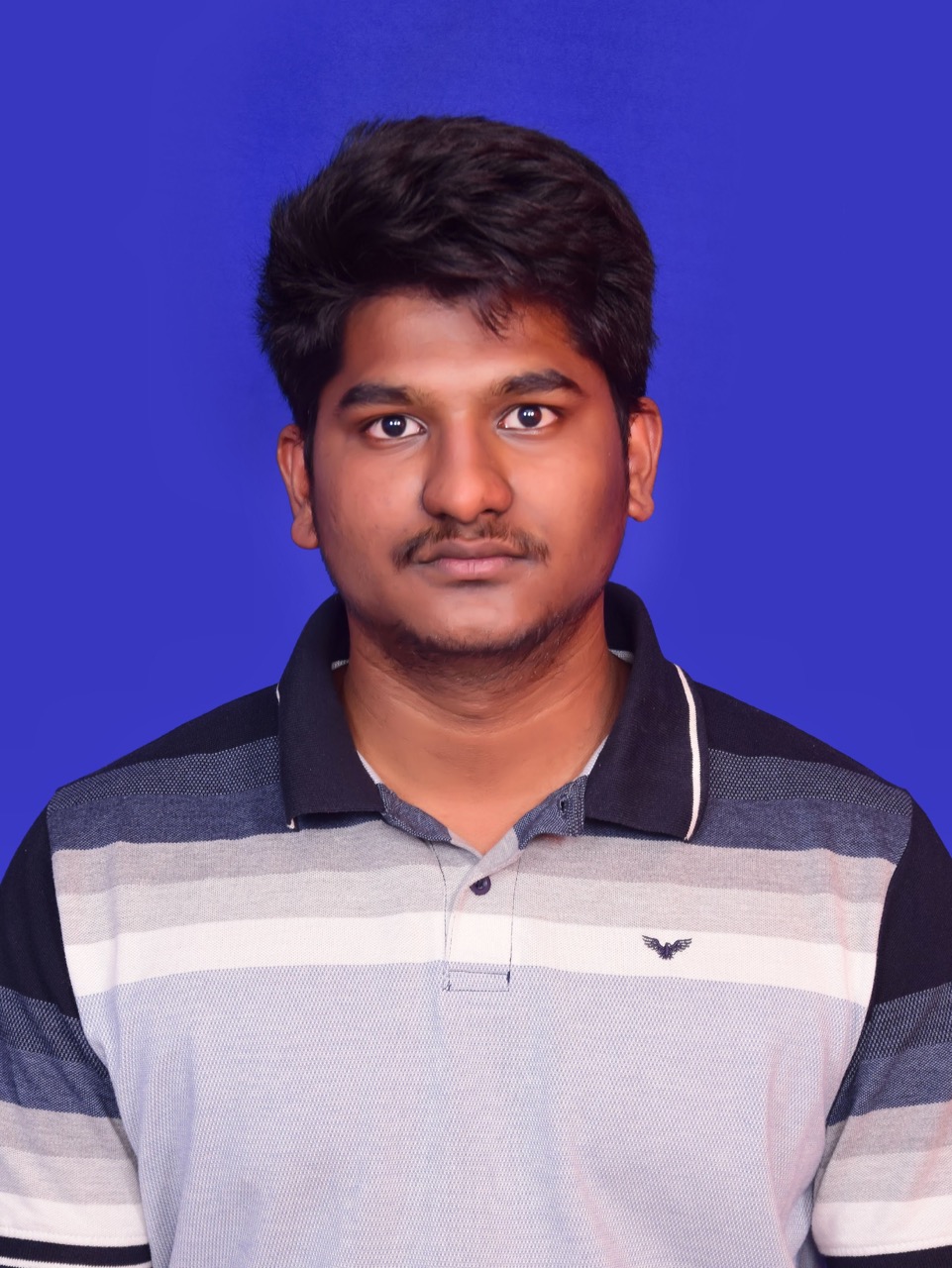}}]{BHARGAV CHIRUMAMILLA~} is currently pursuing the B.Tech in computer science and engineering from the Indian Institute of Technology - Dharwad (IIT Dharwad), India. His research interests include AI/ML, 5G and Beyond Technology.
\end{IEEEbiography}
\vspace{-0.5cm}
\begin{IEEEbiography}[{\includegraphics[width=1in,height=1.25in,clip,keepaspectratio]{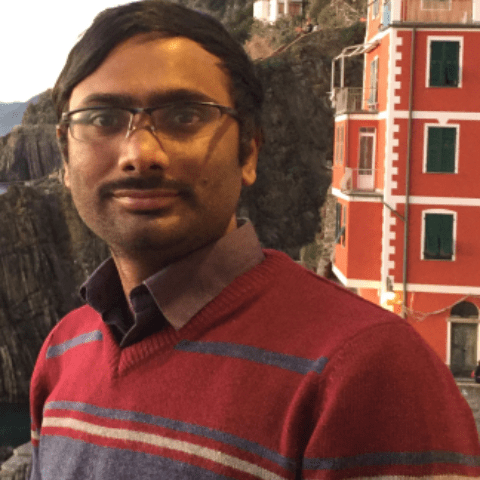}}]{VENKATARAMI REDDY~} (Graduate Student Member, IEEE) received the B.Tech. degree in computer science and engineering from Jawaharlal Nehru Technological University Hyderabad, India, in 2008, and the M.Tech. degree in computer science and engineering (information security) from the National Institute of Technology Calicut, India, in 2010. He is currently an Assistant Professor with the National Institute of Technology Calicut (NITC), India. His research interests include 5G, network function virtualization (NFV), software defined networking, and AI in mobile networks.
\end{IEEEbiography}
\vspace{-0.5cm}
\begin{IEEEbiography}[{\includegraphics[width=1in,height=1.25in,clip,keepaspectratio]{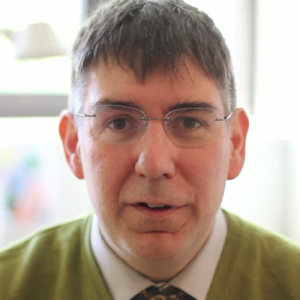}}]{PIERO CASTOLDI~} (Senior Member, IEEE) is currently a Full Professor and the Leader of the ‘‘Networks and Services’’ research area at the TeCIP Institute, Scuola Superiore Sant’Anna, Pisa, Italy. He is also serving as a member of the Management Board of the Consorzio Nazionale Interuniversitario per le Telecomunicazioni (CNIT) and the Executive Board of the Inphotec Foundation, and he is the Director of the Erasmus Mundus Master on Photonic Integrated Circuits, Sensors and Networks (PIXNET). His research interests include telecommunications networks and system both wired and wireless, and more recently reliability, switching paradigms and control of optical networks, including application-network cooperation mechanisms, in particular for cloud networking.
\end{IEEEbiography}
\vspace{-0.5cm}
\begin{IEEEbiography}[{\includegraphics[width=1in,height=1.25in,clip,keepaspectratio]{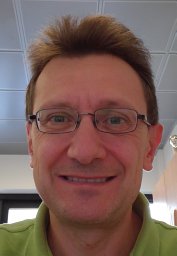}}]{LUCA VALCARENGHI~} (Senior Member, IEEE) has been an Associate Professor with Scuola Superiore Sant’Anna, Pisa, Italy, since 2014. He published almost 300 articles (source Google Scholar, in May 2020) in international journals and conference proceedings. He received a Fulbright Research Scholar Fellowship, in 2009, and a JSPS ‘‘Invitation Fellowship Program for Research in Japan (Long Term),’’ in 2013. His research interests include optical networks design, analysis, and optimization; communication networks reliability; energy efficiency in communications networks, optical access networks, zero touch networks and service management, experiential networked intelligence, and 5G technologies and beyond.
\end{IEEEbiography}
\vspace{-0.5cm}

\begin{IEEEbiography}[{\includegraphics[width=1in,height=1.25in,clip,keepaspectratio]{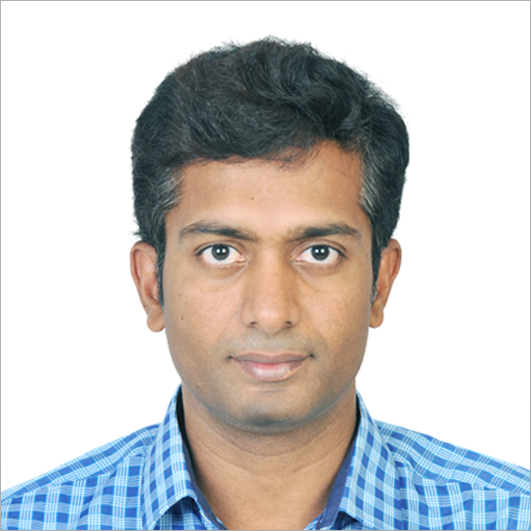}}]{BHEEMARJUNA REDDY TAMMA~} (Senior Member, IEEE) received the Ph.D. degree from the Indian Institute of Technology Madras, India, and then worked as a Postdoctoral Fellow with the University of California at San Diego prior to taking up faculty position with the Indian Institute of Technology Hyderabad. He is a Professor with the Department of CSE, Indian Institute of Technology Hyderabad. His research interests are in the areas of converged cloud radio access networks, SDN/NFV for 5G, network security, and green ICT.
\end{IEEEbiography}
\vspace{-0.5cm}
\begin{IEEEbiography}[{\includegraphics[width=1in,height=1.25in,clip,keepaspectratio]{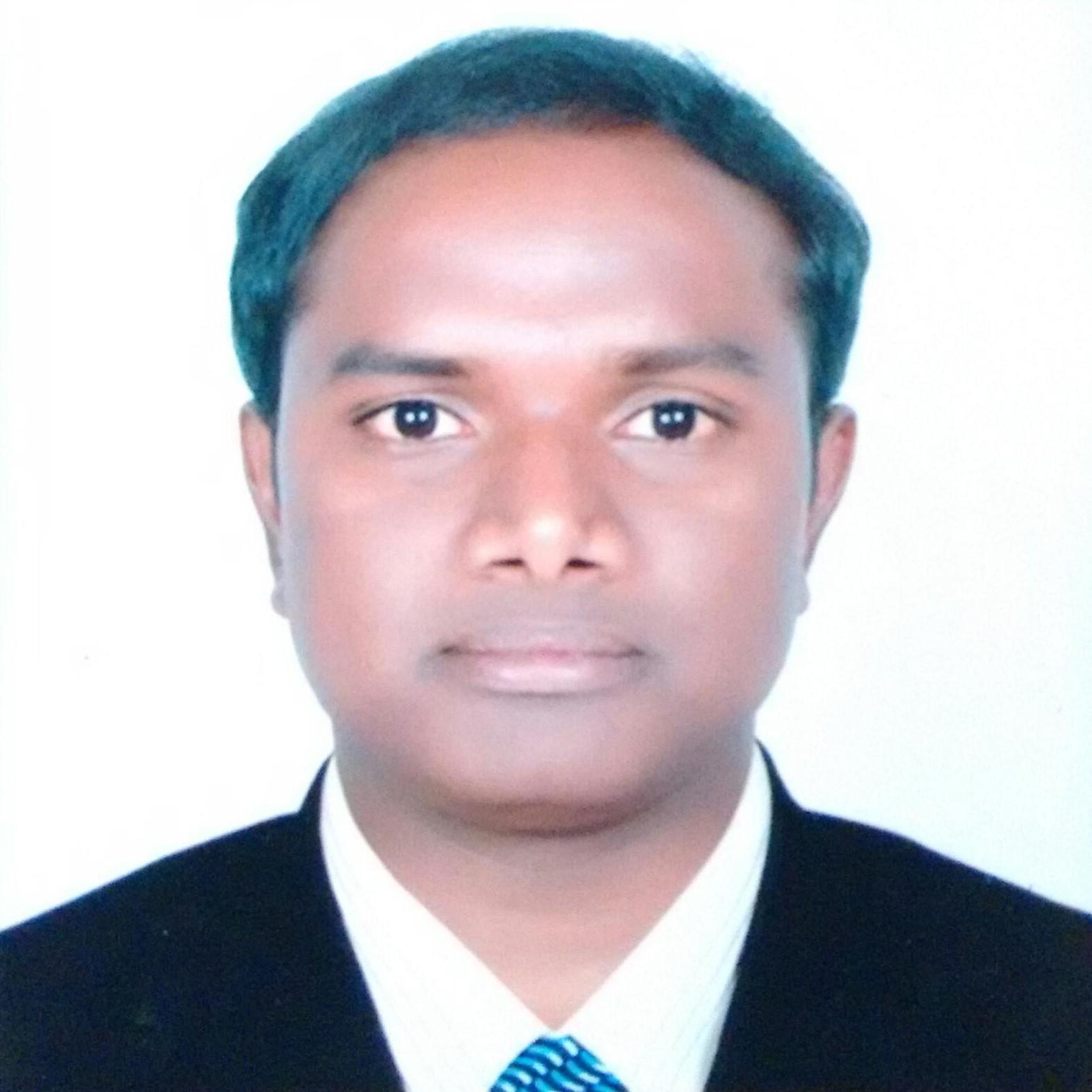}}]{\textbf{KOTESWARARAO KONDEPU}~} (Senior Member, IEEE) received the Ph.D. degree in computer science and engineering from the Institute for Advanced Studies Lucca (IMT), Italy, in July 2012. He is currently an Assistant Professor with the Indian Institute of Technology Dharwad, Dharwad, India. His research interests include 5G, optical networks design, energy-efficient schemes in communication networks, and sparse sensor networks.
\end{IEEEbiography}

\end{document}